\documentclass[prb,a4paper,superscriptaddress,twocolumn]{revtex4-2}
\usepackage{amssymb}
\usepackage{amsmath}
\usepackage{amsfonts}
\usepackage{graphicx}
\usepackage{float}
\usepackage{bm}
\usepackage{color, comment}
\usepackage{hyperref}
\usepackage{comment}

\usepackage{xcolor}

\definecolor{darkGreen}{RGB}{200,32,96}

\def\bQ{\mathbf{Q}}

\def\bp{\mathbf{p}}
\def\pf{p_{\rm F}}
\def\sigmap{\sigma^\prime}
\def\ve{\varepsilon}
\def\vf{\varphi}
\def\vfQ{\varphi_{\mathbf{Q}}}
\def\br{\mathbf{r}}
\def\bQ{\mathbf{Q}}

\begin{document}
\title{Nesting-driven ferromagnetism of itinerant electrons}

\author{Ya.I. Rodionov}
\affiliation{Institute for Theoretical and Applied Electrodynamics, Russian
Academy of Sciences, 125412 Moscow, Russia}
\affiliation{National Research University Higher School of Economics, Moscow 101000, Russia}

\author{A.V. Rozhkov}
\affiliation{Institute for Theoretical and Applied Electrodynamics, Russian
Academy of Sciences, 125412 Moscow, Russia}

\author{M.E.S. Beck}
\affiliation{National University of Science and Technology MISIS, Moscow, 119049 Russia}

\author{A.O. Sboychakov}
\affiliation{Institute for Theoretical and Applied Electrodynamics, Russian
Academy of Sciences, 125412 Moscow, Russia}

\author{K.I. Kugel}
\affiliation{Institute for Theoretical and Applied Electrodynamics, Russian
Academy of Sciences, 125412 Moscow, Russia}
\affiliation{National Research University Higher School of Economics, Moscow 101000, Russia}

\author{A.L. Rakhmanov}
\affiliation{Institute for Theoretical and Applied Electrodynamics, Russian
Academy of Sciences, 125412 Moscow, Russia}

\date{\today}

\begin{abstract}
We theoretically investigate a model with electrons and holes whose Fermi
surfaces are perfectly nested. The fermions are assumed to be interacting,
both with each other and with the lattice. To suppress inhomogeneous
states, a sufficiently strong long-range Coulomb repulsion is included into
the model. Using the mean field approximation, one can demonstrate that in the absence of doping, the ground state of such a model is insulating and possesses a
density-wave order, either SDW, or CDW. Upon doping, a finite ferromagnetic
polarization emerges. It is argued that the mechanism driving the
ferromagnetism is not of the Stoner type.
A phase diagram of the model is constructed, and
various properties of the ordered phases, such as half-metallicity and cone magnetic structure, are studied.

\end{abstract}

\maketitle

\section{Introduction}

Can a gas of itinerant fermions with spin-independent repulsion exhibit
the ferromagnetic state? This is one of the oldest research
topics~\cite{Bloch}
in theoretical condensed matter physics. Typically, the emergence of a finite
ferromagnetic polarization of a Fermi gas is viewed through the lens of the
so-called Stoner criterion. Since the moment of its inception almost one
hundred years ago~\cite{stoner1938collective},
this theoretical device has become a standard tool employed in numerous
original papers, and taught in various many-body-theory textbooks, both
classical~\cite{Huang}, and modern~\cite{blundell2001magnetism}.

The most seductive features of the Stoner criterion are its simplicity and
universality: in the words of Kerson Huang (see p.~274 in
Ref.~\onlinecite{Huang}),
``... if the repulsive strength is sufficiently strong, the system becomes
ferromagnetic.''
For numerous theoretical many-fermion models such a requirement is easy to
benchmark against, and is not too restrictive, turning the criterion into
an accessible tool to study a ferromagnetic instability. One can cite
Refs.~\onlinecite{trilayer2021FraM_nature, 2d_stoner2024Raines,
dft_stoner2017cab6} as recent examples of this approach.

Yet we must weight the simplicity of the Stoner framework against its unclear
reliability. The calculation leading to the Stoner criterion is performed in the paradigm of the perturbation theory in the electron-electron interaction coupling constant. Yet the criterion itself states that the system becomes ferromagnetic when the coupling constant becomes larger than unity, far beyond the applicability of the perturbation theory. This makes the Stoner approach an uncontrollable approximation,
whose
accuracy can be assessed using numerical data only (see, for instance,
Ref.~\onlinecite{no_FM2020holzmann}
and the analysis therein). As we argue in our discussion section below,
there is ample evidence pointing to unreliability of the Stoner-criterion
predictions.

We believe that stable ferromagnetism in a system of itinerant fermions
remains an important research issue of modern theoretical many-body
physics. In this paper, we investigate a specific model where a
ferromagnetic state can arise.

Apart from the Stoner approach, which becomes operational at moderate-to-strong
inter-fermion repulsion,
another mechanism, applicable at arbitrarily weak electron-electron interaction, has been predicted to cause electron gas to become ferromagnetic.
Back in~1975,
Ref.~\onlinecite{volkov1975}
studied the ground state of a weakly interacting electron gas in the framework of Keldysh--Kopaev
Hamiltonian~\cite{keldysh1965}.
The founding block of this model is the existence of nested Fermi surfaces of electrons and holes bound by electron-electron interaction.  The treatment of the paper is phrased in terms of excitonic states. Due to the combined effect of Fermi-surface nesting and electron-electron interaction, the
ferromagnetic
state becomes the most favorable state of the system. The applicability of the discussed mechanism at weak-coupling regime can be viewed as a significant advantage. This implies that, at least in
principle, the accuracy of the derived results can be tested using a suitably
designed perturbative scheme, with the repulsion strength being a small
parameter.  Twenty four years later the experiment\cite{young1999high} corroborated the theoretical predictions outlined in
Refs.~\onlinecite{volkov1975}
and~\onlinecite{keldysh1965}
where a weak ferromagnetism was measured in
Ca$_{1-x}$La$_x$B$_6$
followed by a theoretical explanation~\cite{zhitomir1999}
 presented in the framework of excitonic states.


The goal of this paper is to undertake a more detailed and comprehensive study of different ground states and phases of the metal in the framework of a general model of a compound with nested Fermi surfaces.
Our approach relies on specific ingredients to generate
stable ferromagnetism. Namely, we assume that the system is composed of
both electrons and holes whose Fermi surfaces, as was mentioned above, are nested. The fermions
interact with each other via the Coulomb repulsion. In addition, the
system Hamiltonian includes coupling between the fermions and lattice
distortions at the nesting vector (this coupling is to some extent optional,
since the ferromagnetic polarization can emerge without it).

The main result of
this study is the detailed  demonstration of a stable ferromagnetism in this model
under doping. We also show that the ferromagnetic state is half-metallic,
and, depending on  system's parameters, it possesses either perfect spin
polarization of the Fermi surface states, or the so-called spin-flavor
polarization~\cite{half_met_rozhkov2017prl, Rakh2018}.

The paper is organized as follows. The model itself is formulated in
Sec.~\ref{sec::model}. The ferromagnetism in the spin-density wave phase
at nonzero  doping is discussed in Sec.~\ref{sec::sdw}.
In Sec.~\ref{sec::cdw}, ferromagnetism of doped charge-density wave phase is studied.
Section~\ref{sec::phase_diag} presents the resultant phase diagram.
Section~\ref{sec::discussion} is dedicated to the discussion.

\section{The model}
\label{sec::model}

The electron system with  an inter-particle repulsion and particle-lattice
interaction is represented by the following Hamiltonian
\begin{gather}
\label{ham_init}
 H = H_0 + H_{l} + H_{le} + H_{ee},
\end{gather}
where
$H_0$
is the Hamiltonian describing free electrons and holes,
$H_{l}$
is the lattice deformation energy,
$H_{le}$
is the term describing the interaction of electrons with the lattice and
$H_{ee}$
is the electron-electron interaction.

To correctly construct the Hamiltonian of free electrons
$H_0$, we divide the momentum space into separate sectors related to electrons and holes.
The corresponding field operator $\psi$ is
split into a sum of contributions with energies close to the Fermi energy
\begin{gather}
\label{sep1}
    \psi_\sigma(\br) = \psi_{\sigma,a}(\br) + e^{i\bQ\br}\psi_{\sigma,b}(\br),
\end{gather}
where
${\bf Q}$
is the nesting vector, and partial field operators
$\psi_{\sigma,a}$
and
$\psi_{\sigma,b}$
are smooth on the scale
$2\pi/|{\bf Q}|$.
Decomposition~(\ref{sep1})
engenders related decomposition of the single-fermion Hamiltonian
\begin{gather}
\label{ham0}
    H_0 = \!\sum\limits_{\sigma}\!\int\!\! dV\! \Big[\psi_{\sigma, a}^\dag\ve_a(\bp)\psi_{\sigma, a}+\psi_{\sigma, b}^\dag\ve_b(\bp+\mathbf{Q})\psi_{\sigma, b}\Big].
\end{gather}
\begin{figure}[t!]
\centering
\includegraphics[width=0.4\textwidth]{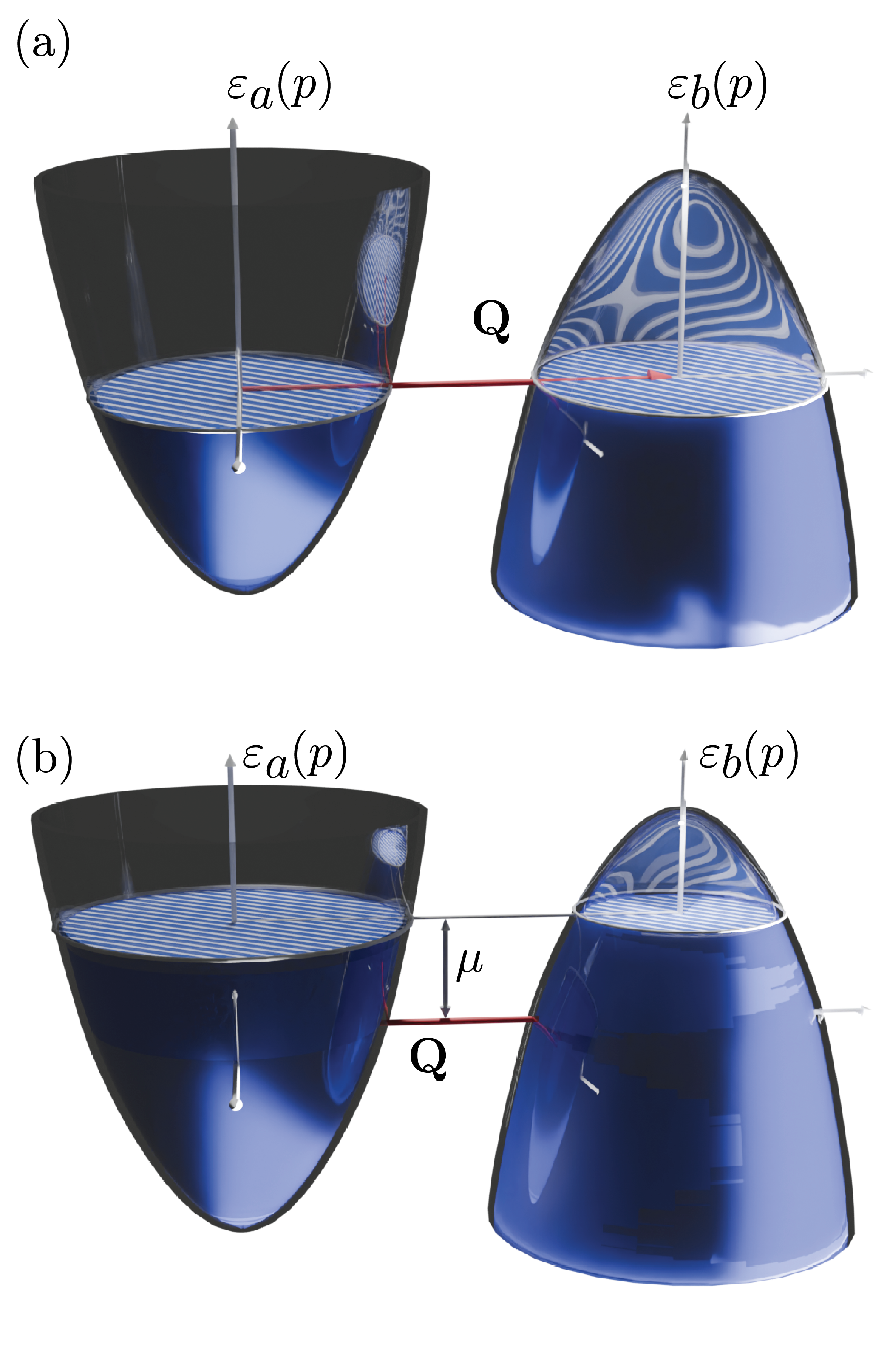}
 \caption{Single-fermion dispersion schematics. (a) The congruence of electron and hole cross sections of  $\ve_{a,b}$ - curves clearly demonstrate the nested  Fermi surfaces separated by the nesting vector $\mathbf{Q}$. (b) The finite doping ($\mu$) destroys the ideal nesting of Fermi surfaces (unequal cross-sections).
\label{fig:spectra0}
}
\end{figure}
In this formula, operators
$\ve_{a,b}$
are defined as follows
\begin{gather}
\label{single_fermion_dispersion}
    \begin{split}
        \ve_a(\bp) &= \frac{\mathbf{p}^2-\pf^2}{2m_a}-\mu,\\
        \ve_b(\bp+\mathbf{Q}) &= -\frac{\mathbf{p}^2-\pf^2}{2m_b}-\mu,
    \end{split}
\end{gather}
where
$\bp = -i\nabla$ (we use the units corresponding to $\hbar = 1$).
The first term in Eq.~(\ref{ham0})
corresponds to electron states, while the second term describes hole
states (see Fig.~\ref{fig:spectra0}). We assume that electrons and holes masses,
$m_a$ and $m_b$, may be unequal $m_a \ne m_b$.
To distinguish between electrons and holes in formulas, it is often
convenient to introduce the so-called fermion-flavor quantum number
$v_\alpha$, where $\alpha = a,b$, and $v_a = 1$,
$v_b = -1$.

If no doping is introduced into the system, the Fermi momenta of both
electrons and holes are identical and equal to
$p_{\rm F}$.
Clearly, this is the case of perfect Fermi surface nesting: momentum-space
translation on
$\bQ$
perfectly superimposes the electron Fermi surface sheet on the hole sheet.
To describe a doped state, we introduce chemical potential $\mu$ in
Eq.~(\ref{single_fermion_dispersion}).
Zero value of $\mu$ corresponds to the undoped state.

The particles interact with each other via the Coulomb repulsion
\begin{eqnarray}
\label{eq::Coulomb}
H^{C}_{ee} = \frac{1}{2}
	\int d\br \int d\br' \frac{e^2}{\epsilon |\br - \br'|}
		\rho (\br) \rho (\br'),
\end{eqnarray}
where $e$ is the elementary charge, and $\epsilon$ is the dielectric constant
of the media. In connection with the density operator
$\rho = \sum_\sigma \psi^\dag_\sigma \psi^{\vphantom{\dag}}_\sigma$,
the following note is in order: due to the field operator
decomposition~\eqref{sep1},
the particle density can be written as a sum of four terms
\begin{eqnarray}
\label{density_decomposition}
\rho (\br) = \rho_a (\br) + \rho_b (\br) + e^{i \bQ \br} \rho_\bQ (\br)
	+ e^{-i \bQ \br} \rho_\bQ^\dag (\br).
\end{eqnarray}
These partial densities are defined as
\begin{eqnarray}
\rho_a = \sum_\sigma
	\psi^\dag_{\sigma,a} \psi^{\vphantom{\dag}}_{\sigma,a},
\quad
\rho_b = \sum_\sigma
	\psi^\dag_{\sigma, b} \psi^{\vphantom{\dag}}_{\sigma, b},
\\
\rho_\bQ^{\vphantom{\dag}} = \sum_\sigma
	\psi^\dag_{\sigma, a} \psi^{\vphantom{\dag}}_{\sigma, b},
\quad
\rho_\bQ^\dag = \sum_\sigma
	\psi^\dag_{\sigma, b} \psi^{\vphantom{\dag}}_{\sigma, a}.
\end{eqnarray}
On the spatial scale $2\pi/|\bQ|$,
their variations are assumed to be insignificant.

As for
$H^C_{ee}$
itself, we must remember that at low energies this ``bare'' interaction is
renormalized by myriads of screening events. Thus, for a low-energy model,
it is reasonable to replace
$H^C_{ee}$
with an effective interaction. It can be cast as a sum of direct and
exchange point-like interactions
\begin{eqnarray}
   H_{ee} = H_{\rm dir} + H_{\rm ex},
\\
\label{ee_dir}
   H_{\rm dir}
=
g\sum\limits_{\sigma\sigma^\prime}
	\int dV \psi^\dag_{\sigma, a}\psi_{\sigma, a}^{\vphantom{\dag}}
		\psi^\dag_{\sigmap, b}\psi_{\sigmap, b}^{\vphantom{\dag}}
		+ \ldots,
\\
H_{\rm ex} =
	g_\bot\sum\limits_{\sigma\sigmap}
	\int dV \psi^\dag_{\sigma, a}\psi_{\sigma, b}^{\vphantom{\dag}}
		\psi^\dag_{\sigmap, b}\psi_{\sigmap, a}^{\vphantom{\dag}} .
\end{eqnarray}
Here, the direct term
$H_{\rm dir}$
represents couplings of the form
$\rho_a \rho_a$,
$\rho_a \rho_b$,
and
$\rho_b \rho_b$.
They describe two-fermion collisions with a small momentum transfer.
Since
$\rho_a \rho_b$
plays a key role in the formation of ordered phases, it is explicitly shown in
Eq.~(\ref{ee_dir}).
The two other coupling types,
$\rho_a \rho_a$
and
$\rho_b \rho_b$,
do not directly affect the ordering.  Thus, we chose to replace them by
ellipsis in
Eq.~(\ref{ee_dir}).

The exchange term
$H_{\rm ex}$
represents
$\rho_\bQ^{\vphantom{\dag}} \rho_\bQ^\dag$
coupling. It describes collisions between an electron and a hole in which
the electron (hole) is scattered to the hole (electron) Fermi surface
sheet. Obviously, such a scattering event transfers large momentum
$\sim \bQ$.
We expect that in a typical situation the effective coupling constants
decrease when the transferred momentum grows. Thus, we have
\begin{eqnarray}
g > g_\bot > 0.
\end{eqnarray}
Both constants are positive due to the inter-particle interaction being
repulsive.

Next, we introduce the Hamiltonian terms responsible for the
electron-lattice coupling. For our analysis, it is sufficient to take the
deformation energy of the lattice in its simplest form, discarding the
shear modulus. Therefore, the Hamiltonian of the distorted lattice reads
\begin{gather}
\label{lattice}
    H_l = \int \frac{K}{2}\varphi^2 dV,
\end{gather}
where
$\varphi = \nabla\cdot\mathbf{u}$
and
$\mathbf{u}$
is the strain vector, and $K$ is the bulk modulus. The
quantum mechanical operator $\vf$ is constructed based on its classical
counterpart via a standard canonical quantization procedure
[\onlinecite{AGD}].

Likewise, the electrons' interaction with the lattice is taken in the standard
form~[\onlinecite{AGD}]
of the electron-phonon interaction:
\begin{gather}
\label{Hle_general}
    H_{le} = \eta \int \rho \varphi dV
= \eta\sum\limits_{\sigma}
		\int \psi_\sigma^\dag\psi_\sigma\varphi dV,
\end{gather}
where $\eta$ is the electron-lattice coupling constant.

It is well-known that a Fermi surface with nesting is unstable with respect
to formation of spin-density wave (SDW), or charge-density wave (CDW)
ordered phases. To account for possible coupling between the CDW order
parameter and electrons, we will be interested in a particular static
distortion of the lattice with the wave vector
$\mathbf{Q}$.
Therefore, instead of considering the full dynamics of the lattice, we
substitute the quantum mechanical operator
$\vf(\br)$
with the classical potential profile
\begin{gather}
\label{lattice_def}
    \vf(\br) = e^{i\mathbf{Q}\br}\vf_\bQ+e^{-i\mathbf{Q}\br}\vf^*_\bQ,
\end{gather}
where
$\vf_\bQ$
and
$\vf_\bQ^*$
represent classical lattice distortion mode with the wave vector
$\bQ$.
Note that, due to the lattice dynamics being much slower than that of
electrons, treating lattice degrees of freedom as classical static
variables is admissible.

Combining the latter expression with
Eqs.~\eqref{density_decomposition}
and~\eqref{Hle_general},
we obtain the following Hamiltonian for lattice and electron-lattice
interaction
\begin{gather}
\label{lat_elec}
    H_l \!+\! H_{le}  = K|\vf_\bQ|^2 V
	\!+\!	
	 \eta \!\left( \!
		\vf^*_\bQ\sum\limits_{\sigma}\!\!
			\int\!\psi^\dag_{\sigma,a}\psi_{\sigma,b}dV
	\!+\! {\rm H.c.}\! \right)\!\!,
\end{gather}
where $V$ stands for the system volume.

\section{Spin density wave}
\label{sec::sdw}

As we already mentioned above, the instability of the nested Fermi surface
in our model leads to the ordering with a density-wave-type order parameter.
Depending on details, the order parameter is either SDW or CDW. The
ferromagnetism we aim to describe in this paper can be viewed as a reaction
of the density wave to doping affecting the gapped ordered environment. To
substantiate this picture, we will use the mean field approximation to study
the two types of density-wave order parameters, both with and without
doping. In this section, we investigate the SDW, starting with the undoped
case.

\subsection{Undoped SDW state}
\label{subsec::sdw_undoped}

To describe the SDW, we assume that the system acquires the nonzero field
expectation value
\begin{gather}
\label{eq::sdw_OP_def}
    \Delta_\sigma = g\langle\psi^\dag_{\sigma,a}
    \psi_{\bar{\sigma}, b}^{\vphantom{\dag}}\rangle.
\end{gather}
Symbol
$\bar \sigma$
in this formula represents the value opposite to $\sigma$. An SDW phase
represented by this order parameter is polarized in $Oxy$ plane.
The corresponding spin polarization in the coordinate space is
\begin{eqnarray}
\label{eq::S_vec}
\left(
	\begin{matrix}
		S_x ({\bf r} )\\
		S_y ({\bf r} )\\
	\end{matrix}
\right)
=
\frac{e^{i {\bf Q} \cdot {\bf r}}}{2g}
\left(
	\begin{matrix}
	 	\Delta_\uparrow + \Delta_\downarrow \\
	 	-i \Delta_\uparrow + i \Delta_\downarrow \\
	\end{matrix}
\right)+ {\rm C.c.}
\end{eqnarray}
Depending on relation between complex order parameters
$\Delta_\uparrow$
and
$\Delta_\downarrow$,
the SDW can be either collinear or coplanar.

The mean field Hamiltonian corresponding to this order parameter reads
\begin{eqnarray}
\label{eq::Hsdw_MF}
H_{\rm sdw}
=
\int \!\!dV\, \sum_\sigma
		\chi^\dag_\sigma
		\mathcal{H}^{\rm sdw}_\sigma
		\chi^{\vphantom{\dag}}_\sigma
+ \frac{V}{g} \sum_\sigma |\Delta_\sigma|^2.
\end{eqnarray}
where the spinor field
$\chi_\sigma$
and a
$2\times 2$
matrix
${\cal H}_\sigma^{\rm sdw}$
are defined according to
\begin{eqnarray}
{\cal H}_\sigma^{\rm sdw}
=
\begin{pmatrix}
	\ve_a & -\Delta_\sigma^* \\
	-\Delta_\sigma  & \ve_b
\end{pmatrix},
\quad
\chi_\sigma
=
\begin{pmatrix}
	\psi_{\sigma,a }\\
	\psi_{\bar{\sigma},b}
\end{pmatrix}.
\end{eqnarray}
Note that, within the mean field approximation, the fermionic fields in the
SDW phase are evidently divided into two fermion sectors: sector
$\sigma$ (sector
$\bar \sigma$) contains
$\psi_{\sigma, a}$
and
$\psi_{\bar{\sigma}, b}$
fields
($\psi_{\bar{\sigma}, a }$
and
$\psi_{{\sigma}, b}$
fields).
Below we will see that this splitting occurs not only in the Hamiltonian, but
in the self-consistency equations as well, and persists at finite doping.
The concept of fermion sectors goes beyond simple notation convenience as
the sectors will play pivotal role in our analysis of the doped system.

Another important observation is that neither
$\vfQ$
nor
$g_\bot$
enter
$H_{\rm sdw}$.
Indeed, the CDW order parameter vanishes in the SDW phase. Thus, at the
mean field level, the value of
$\vfQ$
is also zero, since
$\vfQ$
can couple to the CDW order parameter only, see
Eq.~(\ref{Hle_general}).
Likewise,
$H_{\rm ex}$
describes interaction between two charge-density waves, and does not
contribute to the spin-density channel.

 Diagonalizing ${\cal H}_{\rm sdw}$,
we obtain the following spectrum
\begin{gather}
\label{spectra0}
    \begin{split}
	    E_{\sigma,\pm} ({\bf p}) &=\frac{\ve_a({\bf p})+\ve_b({\bf p+\mathbf{Q}})}{2}\\
        &\pm\sqrt{\Big[\frac{\ve_a({\bf p})-\ve_b({\bf p}+\mathbf{Q})}{2}\Big]^2
			+|\Delta^2_\sigma|}.
    \end{split}
\end{gather}
At zero temperature
$T=0$, all states with
$E_{\sigma,\pm}>0$
are empty, only the states with
$E_{\sigma,\pm}<0$
contribute to the total energy $E$. It is equal to
\begin{gather}
E = \sum\limits_{\sigma,\bp}E_{\sigma,-}(\bp)
+
\frac{V}{g} \sum_\sigma |\Delta_\sigma^2|.
\end{gather}
Differentiating $E$ with respect to
$\Delta_\sigma$,
one can derive the following self-consistency equation
\begin{eqnarray}
\label{min_D}
\Delta_\sigma = g N_{\rm F} \Delta_\sigma \frak{I}_\Lambda (|\Delta_\sigma|).
\end{eqnarray}
In this formula,
$N_{\rm F}$
is the density of states at the Fermi level
\begin{eqnarray}
N_{\rm F} = \frac{\pf m_a m_b}{\pi^2(m_a+m_b)}
=
\frac{\pf m_\Sigma}{4\pi^2}\gamma^2,
\end{eqnarray}
where we introduced the notation:
\begin{gather}
\label{gamma}
    \begin{split}
        \gamma   &=\sqrt{1-\frac{\delta m^2}{m^2_\Sigma}}
                  = \frac{2\sqrt{m_a m_b}}{m_\Sigma}\leq 1,
\\
 m_\Sigma &=m_a+m_b,\quad \delta m = m_b - m_a.
    \end{split}
\end{gather}
As for
$\frak{I}_\Lambda ({\cal E})$
in
Eq.~(\ref{min_D}),
this function, being defined as
\begin{gather}
\label{int}
\frak{I}_\Lambda({\cal E})
=
\int\limits_{0}^\Lambda
	\frac{p^2dp}{4\pi^2 N_{\rm F} }
	\frac{1}{\sqrt{
			\frac{1}{4} [\ve_a ({\bf p}) -\ve_b ({\bf p}) ]^2
			+
			\mathcal{E}^2
		}},
\end{gather}
demonstrates the familiar logarithmic divergence in the limit of small
${\cal E}$
\begin{eqnarray}
\label{aux_int1}
\frak{I}_\Lambda({\cal E})
\approx \ln \left(\frac{\varepsilon_*}{\cal E} \right),
\end{eqnarray}
where
$\ve_*\sim \Lambda \pf/(m_a+m_b)$
is the ultraviolet energy cutoff corresponding to the large momentum cutoff $\Lambda$. Here we note
that the integral~\eqref{aux_int1} as well as all other integrals over
filled states below are evaluated with the help of expansion of the density
of states near the Fermi surface and the subsequent linearization of the
particle-hole spectra $\ve_{a, b}(\mathbf{p})$. The results, therefore, are
insensitive to the shape of the Fermi surface, as usually the case in
Fermi-liquid-like theories.

One can see that
expression~(\ref{min_D})
is a set of two decoupled equations labeled by index $\sigma$. Thus, as we
mentioned before, the self-consistency equations for the two fermionic
sectors are decoupled from each other. This feature is unique to the SDW
mean field formalism, and it drastically simplifies our calculations. This
simplification becomes particularly obvious when doped systems are studied.

Solving
Eq.~(\ref{min_D}),
we discover that, within each sector, there is either the trivial solution
$\Delta_\sigma = 0$,
or
\begin{gather}
\label{eq::bcs_solution}
\Delta_\sigma = \Delta_0 e^{i \Phi_\sigma},
\end{gather}
where the energy scale
\begin{eqnarray}
\label{Delta0_def}
\Delta_0 = \ve_* e^{-\frac{1}{gN_{\rm F}}},
\end{eqnarray}
satisfies a BCS-type equation
\begin{gather}
\label{min3}
\frac{1}{g N_{\rm F}} = \frak{I}_\Lambda(\Delta_0).
\end{gather}
Phases $\Phi_\sigma$ remain undetermined by the
self-consistency equations, and can be chosen arbitrary.

The many-electron state with
$\Delta_\sigma = 0$
is unstable, and the ground state is described by
Eq.~(\ref{eq::bcs_solution})
for both
$\sigma = \uparrow, \downarrow$.
Since the order parameters are finite in both sectors, we conclude that the
mean field SDW phase is insulating as its
spectrum~(\ref{spectra0})
has a gap
\begin{eqnarray}
\label{gap0}
\Delta E = 2 \gamma \Delta_0.
\end{eqnarray}
This gap is not direct: the bottom of the conduction band is not at
$p=p_{\rm F}$,
but rather is at
$p_{\rm c}
=
p_{\rm F} - \gamma \delta m \Delta_0 /(2 p_{\rm F} )$,
while the top of the valence band is at
$p_{\rm v}
=
p_{\rm F} + \gamma \delta m \Delta_0 /(2 p_{\rm F} )$,
and
$p_{\rm c} \ne p_{\rm v}$.
In the limit of equal masses $m_a = m_b$,
this gap becomes direct and equal to
$2 \Delta_0$.

Once we solve the self-consistency equations,  we obtain the standard formula for the ground state energy
\begin{gather}
    \label{ground}
    F_0 = - N_{\rm F} \Delta_0^2 V.
\end{gather}

Next, using Eq.~(\ref{eq::S_vec}),
we calculate SDW spin polarization in coordinate space
\begin{eqnarray}
\label{eq::S_vec_undoped}
\left(
	\begin{matrix}
		S_x ({\bf r} )\\
		S_y ({\bf r} )\\
	\end{matrix}
\right)
=
\frac{2 \Delta_0}{g}
\left(
	\begin{matrix}
		\cos \Phi\\
		\sin \Phi\\
	\end{matrix}
\right)
\cos ( {\bf Q} \cdot {\bf r} + \phi ).
\end{eqnarray}
The value of
$\phi = (\Phi_\sigma+\Phi_{\bar{\sigma}})/2$ determines overall space translation of the SDW texture, while
$\Phi = (\Phi_\sigma-\Phi_{\bar{\sigma}})/2$ sets the direction along which the SDW is polarized. Spin polarization~(\ref{eq::S_vec_undoped})
is explicitly collinear.

The degeneracy with respect to $\Phi$ is a consequence of model's spin-rotation
symmetry. This degeneracy cannot be lifted unless the latter is broken. The
degeneracy relative to $\phi$ may be removed, at least partially, if
lattice effects are introduced in the form of umklapp terms. For example,
if
$2\bQ$
is either zero or a reciprocal lattice vector, umklapp couplings with
structure
$\psi^\dag_{\sigma, a} \psi^{\vphantom{\dag}}_{\sigma, b}
\psi^\dag_{\sigma',a} \psi^{\vphantom{\dag}}_{\sigma', b}$
and
$\psi^\dag_{\sigma, b} \psi^{\vphantom{\dag}}_{\sigma,a}
\psi^\dag_{\sigma', b} \psi^{\vphantom{\dag}}_{\sigma',a}$
become permissible. They ``pin'' the SDW to the lattice, which restricts
possible values of $\phi$ to a finite set.

\subsection{SDW at finite doping}
\label{subsec::sdw_doped}

Adding charge carriers, either electrons or holes, destroys perfect nesting of the
Fermi surface in our system (see Fig.~\ref{fig:spectra0}). The doped state has a tendency toward the
formation of spatially inhomogeneous states, which have been studied in
numerous publications in various
contexts~\cite{gorbatsevich1992band, stripes_pnictides2010gorkov,
	igoshev2010, Igoshev2015hubbard,
nesting_review2017, pressure_phasep_rakhmanov2020, kokanova2021,KaganKugRakhPhRep2021,BookPS}.
Sufficiently strong Coulomb long-range repulsion restores the stability of
a homogeneous phase, causing system to lower its energy by implementing a
kind of phase separation that occurs in space of discrete indices, instead
of continuous coordinate
space~\cite{half_met_rozhkov2017prl, Rakh2018, neutron_half_met2020,
jetp_lett_minireview_half_met2020, fracmet2021prblett, AB_quarter_metal}.
In such a regime interaction effects force uneven distribution of electrons
among available discrete quantum numbers. This unevenness, as we will
demonstrate in this section, is ultimately responsible for stabilization of
a ferromagnetic phase in our model.

On the technical side, to introduce doping into the formalism, we will
assume that the chemical potential is no longer zero. For definiteness,
$\mu > 0$
case is studied below. Adaptation for negative $\mu$ is quite
straightforward. We also assume the doping level to be rather small
\begin{gather}
\label{chemical}
\mu \ll\frac{p_{\rm F}^2}{m_{\Sigma}}
\end{gather}
to be able to capture the important physics attributed to the density wave gap formation.

\begin{figure}[t!]
\centering
\includegraphics[width=0.5\textwidth]{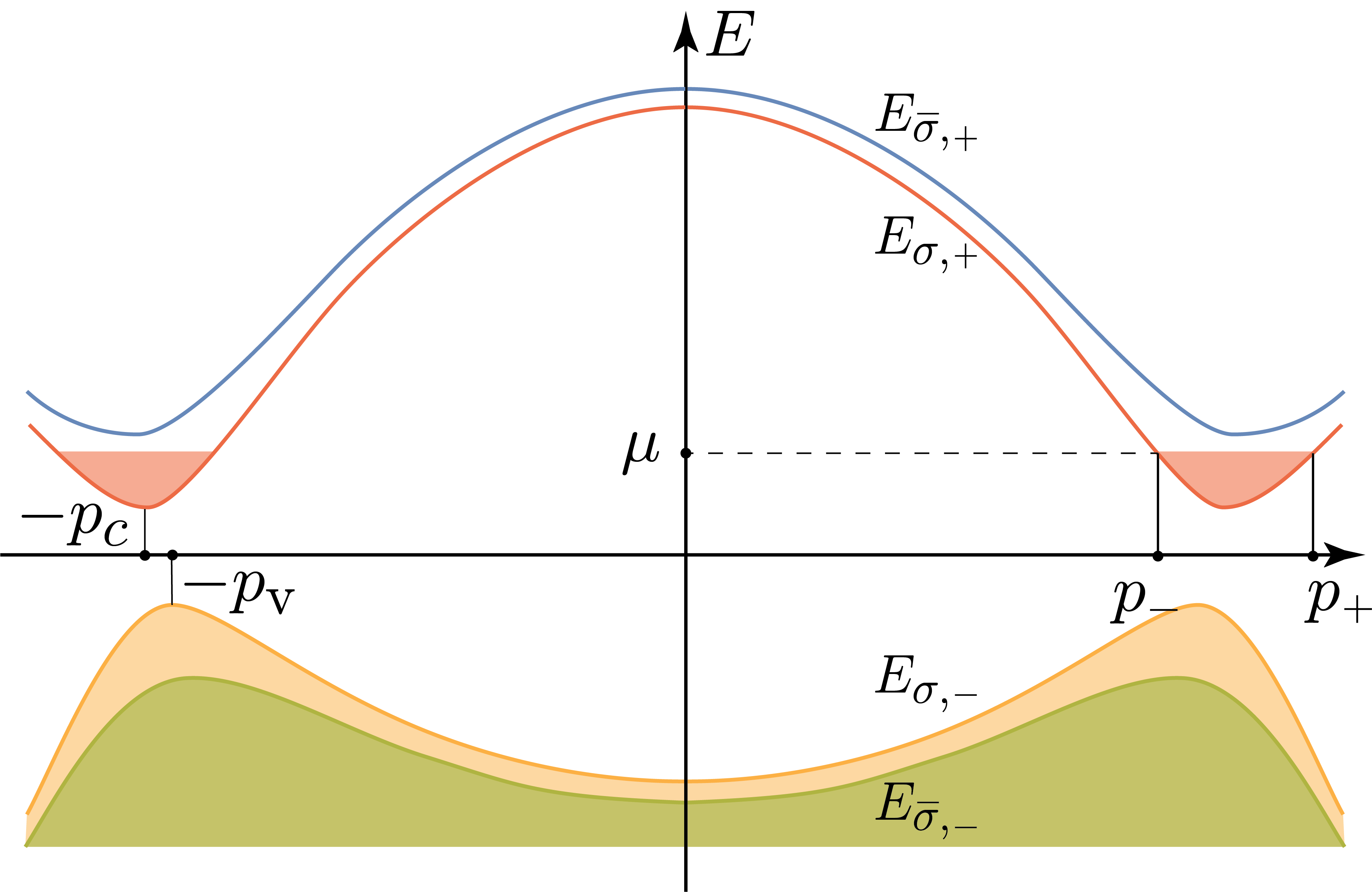}
 \caption{Dispersion curves
$E_{\sigma,\pm}$,
see
Eq.(21),
modified to account for finite doping. The electron mass
$m_a$
is larger than the hole mass
$m_b$.
To simplify the drawing, we assumed that the nesting vector
${\bf Q}$
is zero. Vertical axis represents energy, horizontal axis is momentum $p$.
The chemical potential level is shown by the horizontal dashed line.
Two
sectors are doped differently: one is empty (the corresponding dispersion
curve lies above $\mu$), the other one contains all extra charges
accumulated between
$p_-$
and
$p_+$.
The momentum $p_c$ corresponding to the minimum of the conduction band differs from the momentum $p_v$ corresponding to the maximum of the valence band
marking the indirect gap [see the discussion after Eq.~\eqref{gap0}].
\label{fig:spectra}
}
\end{figure}

Thus, the grand canonical potential $\Omega$ for finite
$\mu$ is a sum
$\Omega = \sum_\sigma \Omega_\sigma$,
where a partial potential for sector $\sigma$ is
\begin{gather}
\Omega_\sigma =
\frac{V}{g} |\Delta_\sigma|^2 +
\sum\limits_{\bp} [E_{\sigma,-}(\bp) - \mu] +
\\
\nonumber
\sum\limits_{\bp}
	[E_{\sigma,+}(\bp) - \mu] \theta (\mu - E_{\sigma,+}(\bp)),
\end{gather}
where
$\theta(x)$
is the Heaviside step-function.
Here, the first sum represents the energy of the filled hole band, while the second sum gives the energy of the \textit{pocket} of the electron band (see Fig.~\ref{fig:spectra}).
For an undoped system,
Eq.~(\ref{eq::bcs_solution})
implies that both fermionic sectors have identical values of
$|\Delta_\sigma|$.
This equivalence may be broken as soon as doping is introduced: it is
argued in
Ref.~\onlinecite{Rakh2018}
that, not only
$|\Delta_\sigma|$
may differ from
$|\Delta_{\bar\sigma}|$,
the doped electrons may be distributed between the sectors unevenly.

To accommodate for this possibility, it is useful to introduce partial
doping concentrations
$x_\sigma = - \partial \Omega_\sigma /\partial \mu$,
whose sum equals to total doping
$x_\sigma + x_{\bar \sigma} = x$.
One can easily establish that
\begin{eqnarray}
\label{eq::partial_doping}
x_\sigma
=
\int\limits_{0}^{+\infty}\frac{p^2dp}{2\pi^2} \theta (\mu - E_{\sigma, +})
\approx
\frac{p_{\rm F}(p_+^2 - p_-^2)}{4\pi^2},
\end{eqnarray}
where the momenta
\begin{eqnarray}
\label{eq::pocket_momenta}
p_\pm
=
\sqrt{\pf^2-\delta m\mu\pm m_\Sigma \sqrt{\mu^2-\gamma^2 |\Delta^2_\sigma|}}
\end{eqnarray}
define two Fermi surfaces in sector~$\sigma$, see
Fig.~\ref{fig:spectra}.
Combining the latter two equations, we ultimately derive
\begin{eqnarray}
\label{eq::partial_doping_final}
x_\sigma
=
\frac{2N_{\rm F}}{\gamma^2}\sqrt{\mu^2-\gamma^2|\Delta_\sigma^2|}.
\end{eqnarray}
Clearly, the square root in this formula is real only when
\begin{eqnarray}
\label{ineq::doped_sector}
\mu > \gamma |\Delta_\sigma|.
\end{eqnarray}
This is the condition for sector $\sigma$ being doped. If this inequality
is violated, the sector is empty of extra charges. The situation, when one
sector is doped, while the other is empty is schematically shown in
Fig.~\ref{fig:spectra}.
We will argue below that, at not too high doping, this corresponds to the
ground state of our model.

For doping $x$ fixed externally, we need to find $\mu$,
$\Delta_\sigma$,
and
$\Delta_{\bar \sigma}$.
Thus, three equations must be determined and then solved. One such equation
is a constraint
\begin{eqnarray}
x =
\frac{2N_{\rm F}}{\gamma^2}
\sum_\sigma \sqrt{\mu^2-\gamma^2|\Delta_\sigma^2|},
\end{eqnarray}
which one obtains using
Eq.~(\ref{eq::partial_doping_final}).
Two minimization conditions
$\partial \Omega/\partial \Delta_\sigma = 0$
complete the desired set of three equations. Performing the
differentiation, one derives the self-consistency equations
\begin{eqnarray}
\label{min2}
\frak{I}_\mu (\Delta_\sigma) = \frac{1}{g N_{\rm F}},
\end{eqnarray}
where function
$\frak{I}_\mu$
is introduced according to
\begin{eqnarray}
\label{eq::I_sigma_def}
\frak{I}_\mu (\Delta_\sigma)
=
\frak{I}_\Lambda (|\Delta_\sigma |)
-
\int\limits_{p_-}^{p_+}
	\frac{p^2dp}{4\pi^2 N_{\rm F} }
	\frac{\theta( \mu - \gamma |\Delta_\sigma| ) }
	{\sqrt{ \Big(\frac{\ve_a-\ve_b}{2}\Big)^2 + |\Delta_\sigma^2|}}.
\end{eqnarray}
Note that, in this definition, the integration limits
$p_\pm = p_\pm (\Delta_\sigma)$
are set by
Eq.~(\ref{eq::pocket_momenta}),
while the integral itself represents the effect of charge carriers inserted
into the system on the system properties. Ultimately, one evaluates
\begin{eqnarray}
\label{eq::I_approx_mu}
\frak{I}_\mu (\Delta_\sigma )
\approx
\ln \left( \frac{\varepsilon_*}{|\Delta_\sigma|} \right)
-
\ln \left(
	\frac{\mu+\sqrt{\mu^2-\gamma^2 |\Delta_\sigma^2|}}{\gamma |\Delta_\sigma|}
	\right)
\end{eqnarray}
in the leading log approximation.
Subtracting
Eq.~\eqref{min3}
from
Eq.~\eqref{min2}
and exploiting the smallness of
$\mu$ (see Eq.~\ref{chemical}),
we obtain the following relation between $\mu$ and
$|\Delta_\sigma|$
\begin{gather}
    \ln \left( \frac{\mu+\sqrt{\mu^2-\gamma^2|\Delta_\sigma^2|}}
	    	{\gamma|\Delta_\sigma|}
	\right)
   -\ln\frac{\Delta_0}{|\Delta_\sigma|} = 0.
\end{gather}
Using
Eq.~(\ref{eq::partial_doping_final})
one can finally establish the relation between
$x_\sigma$ and $\Delta_\sigma$
\begin{eqnarray}
\label{minSol1}
|\Delta_\sigma|
=
\Delta_0\sqrt{1-\frac{x_\sigma\gamma }{N_{\rm F}\Delta_0}}.
\end{eqnarray}
We see that, as anticipated, the SDW sectors remain decoupled even at
finite doping. Also,
$|\Delta_\sigma|$
is a decreasing function of the partial doping: for finite
$x_\sigma$,
one has
$|\Delta_\sigma| < \Delta_0$.

It is not difficult to demonstrate that the chemical potential is a linear
function of the partial doping
\begin{eqnarray}
\label{muSol1}
    \mu = \Delta_0\gamma-\frac{\gamma^2 x_\sigma}{2N_{\rm F}}.
\end{eqnarray}
In addition, one might conclude that this formula immediately implies that
$x_\sigma = x_{\bar{\sigma}}$:
the chemical potential is the same for both sectors, so partial doping concentrations must be equal.

However, the latter reasoning is, in fact, incorrect. Let us remind the
reader that, per our assumption at the beginning of the derivation, both
Eqs.~(\ref{minSol1})
and~(\ref{muSol1})
are valid for finite
$x_\sigma$
only. When
inequality~(\ref{ineq::doped_sector})
is satisfied for sector $\sigma$, but violated for sector $\bar\sigma$,
then all the doping is in sector $\sigma$, while
$x_{\bar\sigma} \equiv 0$.

This suggests that, at finite doping $x$, we have two possibilities:
\begin{eqnarray}
&{\rm Sym.\ case:}& x_\sigma = x_{\bar \sigma} = x/2,
\\
&{\rm Asym.\ case:}& x_\sigma = x, \quad x_{\bar \sigma} = 0.
\end{eqnarray}
These states have non-identical free energies. Indeed, a partial (per
sector) free energy
$F_\sigma = F_\sigma (x_\sigma)$
associated with doping reads
\begin{eqnarray}
\frac{F_\sigma (x_\sigma)}{V}
=
\int_0^{x_\sigma} \mu (x_\sigma') d x_\sigma'
=
\gamma \Delta_0 x_\sigma -\frac{\gamma^2 x_\sigma^2}{4N_{\rm F}}.
\end{eqnarray}
The total free energy
$F = F_0 + F_\sigma + F_{\bar \sigma}$,
where $F_0$ is the free energy of the undoped state (see Eq.~\ref{ground}),
can be expressed as
\begin{eqnarray}
\label{freeEnergy}
\frac{F}{V}
=
-N_{\rm F}\Delta_0^2 + \gamma \Delta_0 x
-\frac{\gamma^2}{4N_{\rm F}}(x_\sigma^2+x_{\bar{\sigma}}^2).
\end{eqnarray}
We therefore have
\begin{eqnarray}
\label{eq::FSymm}
&{\rm Sym.\ case:}& \frac{F_{\rm S}}{V} = - N_{\rm F} \Delta_0^2
	+ \gamma \Delta_0 x -\frac{\gamma^2}{8N_{\rm F}}x^2,
\\
\label{eq::FAsymm}
&{\rm Asym.\ case:}& \frac{F_{\rm A}}{V} = - N_{\rm F} \Delta_0^2
	+ \gamma \Delta_0 x -\frac{\gamma^2}{4N_{\rm F}}x^2.
\end{eqnarray}
Comparing the two free energies, we conclude that asymmetric state, where
all extra electrons are placed into a single sector, corresponds to the
lowest energy. In other words,
Fig.~\ref{fig:spectra}
depicts the state with the lowest energy.

%

\subsection{Properties of the half-metal state}

Characteristic feature of the doped SDW has been discussed in several
publications~\cite{half_met_rozhkov2017prl, Rakh2018,
jetp_lett_minireview_half_met2020, neutron_half_met2020,
fracmet2021prblett, AB_quarter_metal}
in different contexts. Thus, we limit ourselves here to a brief discussion
only. We start with the observation that
Eqs.~(\ref{eq::FSymm})
and~(\ref{eq::FAsymm}),
taken at their face value, clearly reveal instability toward the phase
separation. Indeed, using the relation
$\mu = \partial F/\partial x$,
one can derive
\begin{eqnarray}
&{\rm Sym.\ case:}&
\mu_{\rm S} = \gamma \Delta_0 - \frac{\gamma^2}{4N_{\rm F}}x,
\\
&{\rm Asym.\ case:}&
\mu_{\rm A} = \gamma \Delta_0 - \frac{\gamma^2}{2N_{\rm F}}x.
\end{eqnarray}
Both
$\mu_{\rm S}$
and
$\mu_{\rm A}$
are decreasing functions of doping, which is a signature of the instability
towards formation of spatially inhomogeneous structures (in stable systems
$\partial \mu /\partial x > 0$).

As electrons are charged particles, the possibility for forming spatially
varying distributions of electrons is severely constrained by the long-range
Coulomb interaction~(\ref{eq::Coulomb}).
Below, we will always assume that the long-range forces are sufficiently
strong to guarantee perfect homogeneity of our system.

Second, Eq.~(\ref{minSol1})
indicates that the doping reduces the strength of the SDW order. For
$x > N_{\rm F} \Delta_0/\gamma$
the order parameter in the charge-carrying sector is zero.

Third, it is important to remember that the ground state is a kind of a
metal: there are single-electron states reaching the chemical potential, as
Fig.~\ref{fig:spectra}
illustrates. There are concentric Fermi surface sheets, whose radii are
$p_\pm$.

However, this is not a usual metal, but rather is an example of half-metal:
of four possible types of fermions (electrons/hole, with spin-up/down),
only two of these four types (that is, a half) reach the Fermi level.
Unlike ``classical'' half-metallic
phase~\cite{first_half_met1983},
which demonstrates perfect spin polarization of the Fermi surface, our
asymmetric state is perfectly polarized in terms of the spin-flavor index.

To explain, what this index is, let us define the spin-flavor operator as
\begin{eqnarray}
\label{eq::S_sf_def}
S_{\rm sf}
=
\sum_{\sigma, \alpha} \sigma v_\alpha
	\int dV \psi^\dag_{\sigma, \alpha}
		\psi^{\vphantom{\dag}}_{\sigma, \alpha} ,
\end{eqnarray}
where
$v_\alpha$
is the fermion-flavor quantum number, see
Sec.~\ref{sec::model}.
The presence of
$v_\alpha$
in
Eq.~(\ref{eq::S_sf_def})
distinguishes
$S_{\rm sf}$
from the familiar $z$-axis spin-projection operator
\begin{eqnarray}
\label{eq::Sz_def}
S_z
=
\sum_{\sigma, \alpha} \sigma
	\int dV \psi^\dag_{\sigma, \alpha}
		\psi^{\vphantom{\dag}}_{\sigma, \alpha} ,
\end{eqnarray}
The field operators satisfy obvious commutation rules
$[ S_{z}, \psi_{\sigma, \alpha } ] = -\sigma \psi_{\sigma, \alpha}$,
and
$[ S_{\rm {sf}}, \psi_{\sigma, \alpha} ]
=
-\sigma v_\alpha \psi_{\sigma, \alpha}$.
Therefore, in addition to the spin quantum number $\sigma$, a field
$\psi_{\sigma, \alpha}$
can be characterized by the spin-flavor projection $\sigma v_\alpha$.

It is easy to check that in the sector $\sigma$, both
$\psi_{\sigma, a}$
and
$\psi_{{\bar\sigma}, b}$
carry the same spin-flavor quantum equal to
$+\sigma$.
In the sector
${\bar \sigma}$,
the field operators correspond to a
${-\sigma}$
quantum of
$S_{\rm sf}$.
That is, the Fermi surface of the doped system is characterized by the
single projection of the spin-flavor operator. The Fermi surface sheet with
the opposite projection of
$S_{\rm sf}$
is absent, since the sector
${\bar \sigma}$
is gapped. Thus, due to the Fermi surface polarization in terms of spin-flavor
index, the doped system can be referred to as
a spin-flavor half-metal (SFHM).

\subsection{The ferromagnetic polarization of the half-metal phase}

Simple calculations~\cite{half_met_rozhkov2017prl}
allow one to prove that total spin-flavor polarization of the SFHM state is
proportional to the doping
$\langle S_{\rm sf} \rangle = \sigma x V$.
It is perhaps less obvious, but more important from a fundamental
perspective, that this state possesses finite spin polarization
$\langle S_z \rangle \ne 0$.
Within the mean field approximation,
\begin{eqnarray}
\label{spin01}
\langle S_z \rangle
     \!=\frac{\sigma}{2}\Bigg[ \!\!\int\limits_{p_-}^{p_+}\frac{p^2dp}{2\pi^2}s(|\Delta_\sigma|)
    -\!\! \int\limits_{0}^\infty \frac{p^2dp}{2\pi^2} s(|\Delta_\sigma|)
    \notag
    \\
	+\!\! \int\limits_{0}^\infty \frac{p^2dp}{2\pi^2} s(\Delta_0)\Bigg],
\quad
\end{eqnarray}
where function
$s = s({\cal E})$
is defined as
\begin{eqnarray}
s({\cal E})
=
\frac{\ve_a-\ve_b}{\sqrt{\Big(\ve_a-\ve_b\Big)^2 + 4{\cal E}^2}}.
\end{eqnarray}
The first term in the right-hand side of
Eq.~\eqref{spin01}
represents the contribution of the doped conduction band in sector
$\sigma$. The second and the third terms correspond to contributions of
completely filled valence bands in sectors $\sigma$ and
$\bar \sigma$.
These two terms sum up into the expression that is small, of the order of
$\pf^3\big[\mathcal{O}(a^2\ln a)\big]$,
where
$a = |\Delta_\sigma| m_\Sigma/\pf^2 \ll 1$
is a small parameter. The contribution due to the doped conduction band,
however, produces the answer of the order of
$\pf^3 \mathcal{O}(a)$.
It is easily computed, and we have the following expression for the spin
polarization of SDW in the leading approximation
\begin{gather}
\label{eq::Sz_sdw_FM}
    \langle S_z \rangle \approx -\sigma  \frac{\delta m}{2m_\Sigma}x.
\end{gather}
We see that this expression is finite only when the symmetry between
electrons and holes is broken
$\delta m \ne 0$.
Otherwise,
$\langle S_z \rangle$
is zero.

Beside $\langle S_z \rangle$,
the $Oxy$
polarization~(\ref{eq::S_vec})
remains nonzero. Using the latter formula, we derive

\begin{eqnarray}
\left(
	\begin{matrix}
		S_x \\
		S_y \\
	\end{matrix}
\right)
=
\frac{1}{g}
\left(
	\begin{matrix}
		|\Delta_\uparrow| \cos (\mathbf{Qr}+\Phi_\uparrow)
		+ 	
		|\Delta_\downarrow| \cos(\mathbf{Qr}+\Phi_\downarrow)\\
		|\Delta_\uparrow| \sin (\mathbf{Qr}+\Phi_\uparrow)
		- 	
		|\Delta_\downarrow| \sin (\mathbf{Qr}+\Phi_\downarrow)\\
	\end{matrix}
\right).
\end{eqnarray}
To make this formula more transparent, let us introduce a complex quantity
${\cal S}({\bf r}) = S_x + i S_y$.
It reads
\begin{eqnarray}
{\cal S}
=
(|\Delta_\uparrow| e^{ 2 i\bQ{\bf r} + 2i\Phi} + |\Delta_\downarrow|)
e^{-i \bQ{\br} - i\Phi_\downarrow}.
\end{eqnarray}
Since the order parameters are not identical,
${\cal S}$
never vanishes. This implies that the $xy$ polarization is not collinear
any more.

\begin{figure}[t!]
\centering
\includegraphics[width=0.5\textwidth]{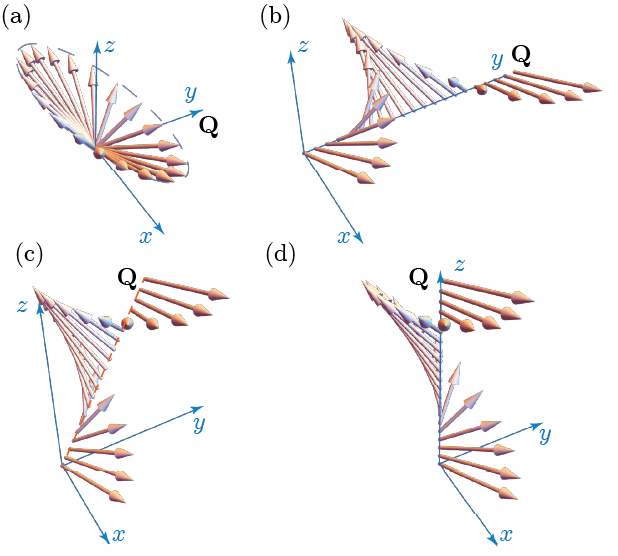}
 \caption{Magnetization vector behavior for different orientation of nesting vector $\bQ$ as radius vector $\br$ moves along $\bQ$. For illustrative purposes, we choose $|\Delta_\downarrow|/ |\Delta_\uparrow|=2$ and  $S_z = 1.3 |\Delta_\uparrow|/g$.
 (a) The magnetic polarization vectors (laid off from the same point) sweeping out a conical shape, as  the radius vector $\br$ moves along the nesting vector $\bQ$. (b-d): the behavior of magnetization vector in coordinate space.
(b)  $\bQ$ is oriented in $y$ direction. (c) $\bQ$ direction is shown with the dashed line, $\bQ \propto [0, 1/\sqrt{5},2/\sqrt{5}]$. (d) $\bQ \propto [0, 0, 1]$, the helical structure is clearly pronounced.
\label{fig:mag}
}
\end{figure}

Assuming for definiteness that
$|\Delta_\uparrow| < |\Delta_\downarrow|$,
one can check trivially that, as $\bQ \br$ changes from zero to $2\pi$,
function
${\cal S}$
winds once around the complex plane origin. This means that local spin
polarization vector
\begin{eqnarray}
{\bf S} ({\bf r}) = (S_x ({\bf r}), S_y ({\bf r}), S_z)^T
\end{eqnarray}
sweeps out a cone as the observation point
${\bf r}$
moves along the nesting vector
$\bQ$ (see Fig.~\ref{fig:mag}).

\section{Charge density wave}
\label{sec::cdw}

In this section, we investigate the doped CDW phase using the mean field
approximation. Dissimilar to the SDW, both exchange and electron-lattice
interactions explicitly enter the CDW calculations. While these
contributions introduce additional complications to theoretical framework,
many features are shared between the SDW and the CDW formalisms.

The CDW corresponds to a different order parameter
\begin{eqnarray}
\label{cdw}
\Delta\equiv\sum\limits_\sigma \Delta_\sigma,
\quad\text{where}\quad
\Delta_\sigma^{\vphantom{\dag}}
=
g\langle\psi^\dag_{\sigma, a}\psi_{\sigma, b}^{\vphantom{\dag}} \rangle.
\end{eqnarray}
It represents spatially modulated density of charge carriers
\begin{eqnarray}
\label{eq::rho_cdw}
\rho ({\bf r})
\!\propto\!
(\Delta_\sigma \!+\! \Delta_{\bar \sigma}) e^{ - i {\bf Q} \cdot {\bf r}}
\!+
{\rm C.c.}
=\! 2 |\Delta| \cos ({\bf Q} \!\cdot\! {\bf r}).
\end{eqnarray}

Here, the origin is assumed to be chosen in a way to eliminate the additional constant phase in the argument of the cosine.
This order parameter is affected by both interaction terms,
$H_{\rm dir}$
and
$H_{\rm ex}$:
the former enhances the CDW, the latter suppresses it. Additionally, it couples
to the lattice distortion
$\vfQ$,
which acts as a stabilization factor for the CDW.

The mean field Hamiltonian for the CDW state can be expressed as
\begin{eqnarray}
\label{Hcdw}
H_{\rm cdw}
=
\int \!\!dV\,
	\sum_\sigma
		\zeta^\dag_\sigma
		{\cal H}_\sigma^{\rm cdw}
		\zeta^{\vphantom{\dag}}_\sigma
	+ V \frak{E}_0^{\rm cdw},
\end{eqnarray}
where a $2\times 2$ matrix
${\cal H}_\sigma^{\rm cdw}$
and spinor field
$\zeta_\sigma$
are equal to
\begin{eqnarray}
\label{eq::H_sigma_def}
{\cal H}_\sigma^{\rm cdw}
=
\begin{pmatrix}
	\ve_a & -D_\sigma^* \\
	-D_\sigma  & \ve_b
\end{pmatrix},
\quad
\zeta_\sigma =
\begin{pmatrix}
	\psi_{\sigma, a}\\
	\psi_{\sigma, b}
\end{pmatrix} ,
\end{eqnarray}
where
\begin{eqnarray}
\label{eq::D_cdw_def}
    D_\sigma = \Delta_\sigma-\frac{g_\bot}{g}\Delta-\eta\vfQ,
\end{eqnarray}
Finally, a $c$-number constant
\begin{eqnarray}
\label{eq::E_cdw_def}
\frak{E}_0^{\rm cdw}
=
- \frac{g_\bot }{g^2}|\Delta|^2
	+
	\frac{1}{g}\sum\limits_\sigma|\Delta_\sigma|^2 + K\vfQ^2
\end{eqnarray}
is a contribution to the total CDW energy density, which emerges in the course of the
mean field decoupling procedure.

Observe that, similar to
$H_{\rm sdw}$,
operator
$H_{\rm cdw}$
splits into two disconnected sectors, see
Eq.~(\ref{Hcdw}).
As in the previous section, these sectors are labeled by index $\sigma$.
However, there are two important differences. First, one can trivially
check that
$\chi_\sigma \ne \zeta_\sigma$.
In other words, the sector content of the SDW phase is not the same as the
sector content of the CDW. Specifically, a CDW sector contains fermions of
identical spin projections. This is not the case for the SDW sectors. Below,
we will see that this CDW feature strengthens the ferromagnetic
magnetization of the doped system.

Second, unlike the SDW case, the separation between the CDW sectors is not
complete. While the mean field Hamiltonian
$H_{\rm cdw}$
is indeed block-diagonal, the inter-sector coupling is present in the
following form: the mean field parameter
$D_\sigma$
in sector $\sigma$ explicitly depends on
$\Delta_{\bar \sigma}$,
which is a characteristics of sector
$\bar{\sigma}$.
As a result, the set of self-consistency equations, in general, does not decouple into two unrelated equations. We will see below that because of this feature, the analysis of the CDW self-consistency conditions is more complicated, and its phase diagram differs from that of the SDW.

The self-consistency equations for the CDW state are derived in a identical to the SDW case manner.
They are
\begin{gather}
\label{eq::cdw_self_cons1}
\frac{ \Delta_\sigma}{g } = N_{\rm F} D_\sigma \frak{I}_\mu (D_\sigma),
\\
\label{eq::cdw_self_cons2}
-K\vf_\bQ
=
\eta N_{\rm F} \sum\limits_\sigma D_\sigma \frak{I}_\mu (D_\sigma),
\end{gather}
where function
$\frak{I}_\mu$
is defined by
Eq.~(\ref{eq::I_sigma_def}).

The lattice strain
$\vf_\bQ$
can be eliminated from the self-consistency system. Indeed, using
Eqs.~(\ref{eq::cdw_self_cons1})
and~(\ref{eq::cdw_self_cons2}),
one can prove that
\begin{gather}
\label{eq::vfQ}
\vf_\bQ = -\frac{\eta\Delta}{Kg}.
\end{gather}
Substituting this into
Eq.~(\ref{eq::D_cdw_def})
we establish
\begin{eqnarray}
\label{eq::D_cdw_effective}
D_\sigma = \Delta_\sigma-\frac{\bar g_\bot}{g}\Delta,
\end{eqnarray}
where effective exchange coupling constant is
\begin{eqnarray}
\label{eq::g_effective}
\bar{g}_\bot = g_\bot - \frac{\eta^2}{K}.
\end{eqnarray}
Inverting
Eq.~(\ref{eq::D_cdw_effective}),
it is possible to prove that
\begin{eqnarray}
\label{eq::Delta_vs_D}
\Delta_\sigma
=
D_\sigma
+
\frac{\bar{g}_\bot}{g - 2\bar{g}_\bot} (D_\sigma + D_{\bar{\sigma}}).
\end{eqnarray}
This allows one to eliminate
$\Delta_\sigma$
in
Eq.~(\ref{eq::cdw_self_cons1})
to obtain
\begin{eqnarray}
\label{eq::self_cons_cdw_final}
\bar{g}_\bot D_{\bar\sigma}
=
N_{\rm F} g (g - 2 \bar{g}_\bot) D_\sigma \frak{I}_\mu (D_\sigma)
+
(\bar{g}_\bot - g) D_\sigma.
\end{eqnarray}
This self-consistency condition is a set of two equations,
labeled by
$\sigma = \uparrow, \downarrow$,
for two unknown variables
$D_\sigma$.

It is not difficult to check that, in the limit
$\bar{g}_\bot = 0$, system~(\ref{eq::self_cons_cdw_final})
decouples into two independent equations. Each equation is mathematically
equivalent to Eq.~(\ref{min2}),
and each defines an order parameter in one sector. To emphasize this
feature, let us rewrite it as follows
\begin{eqnarray}
\label{eq::self_cons_cdw_W}
W D_{\bar \sigma}
=
D_\sigma \left[ \frak{I}_\mu (D_\sigma) - \frac{1}{g N_{\rm F}} - W
\right],
\end{eqnarray}
where the dimensionless constant $W$ is
\begin{eqnarray}
\label{eqq::W_def}
W = \frac{1}{gN_{\rm F}}\frac{\bar{g}_\bot}{g - 2\bar{g}_\bot}.
\end{eqnarray}
Clearly, $W$ vanishes when
$\bar{g}_\bot$
vanishes. For
$W=0$
the decoupling of
system~(\ref{eq::self_cons_cdw_W})
becomes explicit, and the individual equations coincide with
Eq.~(\ref{min2}).
This form of the self-consistency equations can be useful for finding the
order parameters in the limit of small $W$.

Equation~(\ref{eq::g_effective})
demonstrates that the electron-lattice interaction enters the mean field
self-consistency equation via the renormalization of the exchange
coupling constant. Such a property, however, is not limited to the
self-consistency system, but exhibits itself in the model's
thermodynamics as well. Namely, using
Eqs.~(\ref{eq::vfQ})
and~(\ref{eq::Delta_vs_D}),
one can derive
\begin{eqnarray}
\label{eq::E_cdw_vs_D}
\frak{E}_0^{\rm cdw}
=
\frac{1}{g} \sum_\sigma D_\sigma^2 + W N_{\rm F} D^2,
\quad
D = \sum_\sigma D_\sigma.
\end{eqnarray}
This form of
$\frak{E}_0^{\rm cdw}$
depends on the renormalized
$\bar{g}_\bot$, rather than on bare
$g_\bot$,
see
definition~(\ref{eqq::W_def}).
Thus, all lattice properties are conveniently hidden in
$\bar{g}_\bot$.
This renormalization has an important consequence: if
the electron-lattice
coupling is sufficiently strong and the lattice is sufficiently soft,
$\bar g_\bot$
can become negative despite
$g_\bot$
being positive. We will see that the sign of
$\bar g_\bot$ manifests itself in the
shift of the relative stability of the SDW and CDW phases.

Additionally, the three equations, namely,
Eqs.~(\ref{eq::H_sigma_def}), (\ref{eq::self_cons_cdw_W}),
and~(\ref{eq::E_cdw_vs_D}),
are sufficient to express the whole mean field framework in terms of
$D_\sigma$
only. This makes mathematical exploration of the resultant theory a much
more straightforward endeavor. As for $\vfQ$ and $\Delta$'s, they can be
recovered once $D$'s are known.

\subsection{Undoped CDW state}
\label{subsec::cdw_zero_doping}

We solve Eq.~(\ref{eq::self_cons_cdw_W})
in various regimes. Similarly to the case of SDW, we start with the discussion of
the undoped CDW state, which corresponds to
$\mu = 0$.
In this situation, one can rewrite
Eq.~(\ref{eq::self_cons_cdw_W})
in a more explicit form
\begin{eqnarray}
\label{eq::self_cons_cdw_W_undoped}
W D_{\bar \sigma}
=
D_\sigma \ln \left( \frac{D_0}{|D_\sigma|} \right) + W D_\sigma,
\end{eqnarray}
where
\begin{eqnarray}
\label{gap_main_CDW}
D_0 = \Delta_0 e^{-2W}
=
\ve_* \exp\left[-  \frac{1}{N_{\rm F}(g - 2\bar{g}_\bot)}\right].
\end{eqnarray}
There are four types of solutions for
Eq.~(\ref{eq::self_cons_cdw_W_undoped}).
Obviously, (i)~there is a trivial solution
$D_\sigma = D_{\bar \sigma} = 0$.
Solution~(ii) corresponds to the ansatz
$D_\sigma = - D_{\bar \sigma} \ne 0$,
for which one obtains
$|D_\sigma| = \Delta_0$.
Next, solution~(iii) is
\begin{eqnarray}
	D_\sigma = D_{\bar \sigma} \equiv D_0.
\end{eqnarray}
Finally, solution~(iv) is characterized by the requirement that, in the
limit
$W \rightarrow 0$,
it becomes
$|D_\sigma| = \Delta_0$,
$|D_{\bar \sigma}| = 0$.
[To justify case~(iv) one must recall that, at
$W=0$,
the choice
$D_\sigma > 0$,
$D_{\bar \sigma} = 0$
is a valid ansatz for
Eq.~\eqref{eq::self_cons_cdw_W_undoped}.
In the limit of small, but finite $W$, a perturbation theory can be used to
learn how does the latter solution evolves at finite inter-sector
coupling.]

Among these four types, the trivial solution will be ignored, since it
corresponds to a maximum of total energy. Solution~(iv) is of little interest
as well: it corresponds to a saddle point of the energy. (At $W=0$,
the latter statement is quite obvious; when $W$ is finite, a continuity
argument stating that the saddle-point structure persists due to the
extremum Hessian being a continuous function of $W$ can be invoked.)

Solution~(ii) belongs to
subsection~\ref{subsec::sdw_undoped},
since it describes the SDW order. Indeed, the CDW order
parameter~(\ref{eq::rho_cdw})
is identical zero, a consequence of
$D_\sigma = - D_{\bar \sigma}$.
At the same time, the $z$-axis local spin polarization
$\langle S_z \rangle \propto D_\uparrow - D_\downarrow$
is finite. Also, solution~(ii) corresponds to the same energy as the SDW state described
in subsection~\ref{subsec::sdw_undoped}.
In fact, all these SDW states can be related to each other by a suitable
spin-rotation transformation.

Solution~(iii) is the desired CDW state, with finite charge
modulation~(\ref{eq::rho_cdw}).
Its single-electron properties can be studied in a manner similar to what
has been done for the SDW phase in
subsection~\ref{subsec::sdw_undoped}.

It is straightforward to obtain the analogue of the formula~\eqref{ground})  for CDW ground state energy. Using~\eqref{Hcdw} and~\eqref{eq::E_cdw_def}, as well as~\eqref{gap_main_CDW}, we obtain:
\begin{gather}
\label{ground2}
    F_0 = -N_{\rm F} D_0^2 V\equiv -N_{\rm F}\Delta_0^2 e^{-4W}.
\end{gather}

Comparing~\eqref{ground} and ~\eqref{ground2}, we see that, when
$\bar{g}_\bot = 0$ ($W=0$),
the mean field energies of the CDW and the SDW coincide. Otherwise, as elucidated in see
Eq.~(\ref{eq::E_cdw_vs_D}) (as well as in~\eqref{ground2}), the energy of the CDW varies while the SDW energy stays the same.
At
$\bar{g}_\bot < 0 \Leftrightarrow W<0)$,
or, equivalently, when
$g_\bot < \eta^2/K$, Eq.~\eqref{ground2} tells us that
the CDW is a ground state of the model.
We see that the electron-lattice
coupling acts to stabilize the CDW phase.

With no lattice, or when the lattice is too stiff, the SDW always
overtakes the CDW due to finite exchange interaction. Specifically, in the
interval
$0 < \bar{g}_\bot < g/2$,
the CDW is a metastable phase, with the SDW being the ground state. In the
limit
$\bar{g}_\bot \rightarrow g/2 - 0$,
the CDW order parameter vanishes, as seen from~\eqref{gap_main_CDW}. When
$\bar{g}_\bot > g/2$,
even metastable CDW phase disappears.

\subsection{CDW at finite doping}
\label{subsec::doped_cdw}

Now, the doped CDW state can be investigated. Since the CDW is the ground state
of the model when $W$ is negative, we derive the self-consistency equations
at finite chemical potential and
$W = - |W| < 0$.
We substitute
expression~(\ref{eq::I_approx_mu})
for
$\frak{I}_\mu$
into
Eq.~(\ref{eq::self_cons_cdw_W})
and obtain
\begin{eqnarray}
\label{main_gap}
|W| \delta_{\bar \sigma}
=
\delta_\sigma \ln[\nu+\sqrt{\nu^2-\delta_{\sigma}^2}] + |W| \delta_\sigma,
\end{eqnarray}
where we introduced two dimensionless parameters
\begin{eqnarray}
\delta_\sigma = \frac{D_\sigma}{D_0},
\quad
\nu = \frac{\mu}{\gamma D_0}.
\end{eqnarray}
The dimensionless chemical potential $\nu$ is confined within the
$0 \leqslant \nu \leqslant 1$ range.  Quantity $\delta_\sigma$ plays the role of the dimensionless order parameter.
As a consistency check, we observe that, if
$\nu = \delta_\sigma \leqslant 1$,
that is, $\nu$ touches the bottom of conductance band in sector $\sigma$,
then this sector is undoped, and
Eq.~(\ref{main_gap})
is equivalent to
Eq.~(\ref{eq::self_cons_cdw_W_undoped}).

Next, we use expression~\eqref{eq::partial_doping_final}
for the partial doping. To make it valid in the CDW phase, one needs to
replace
$\Delta_\sigma$
with
$D_\sigma$.
Defining dimensionless partial doping
\begin{eqnarray}
\label{eq::Xsigma}
     X_\sigma = \frac{\gamma x_\sigma}{2N_{\rm F}D_0},
\end{eqnarray}
we write
\begin{gather}
\label{connect}
     X_\sigma^2 =
     \nu^2-\delta_\sigma^2.
\end{gather}
Removing $\nu$ from
Eq.~(\ref{main_gap}) by plugging in Eq.~\eqref{connect}, we obtain
the following form of the self-consistency condition
\begin{eqnarray}
\label{self1}
\delta_\sigma \ln[X_\sigma +\sqrt{X_\sigma^2+\delta_\sigma^2}] =
|W|\Big(\delta_{\bar\sigma} - \delta_\sigma \Big).
\end{eqnarray}
As in
subsection~\ref{subsec::sdw_doped},
for fixed total doping
$X = X_\sigma + X_{\bar \sigma}$,
we must find all possible solutions of the self-consistency equations, and
choose the one with the lowest energy.

Assuming that both sectors are equally doped,
$X_\sigma = X_{\bar\sigma} = X/2$,
we derive
\begin{eqnarray}
\label{self_cdw_equal}
\delta_\sigma
\ln \left(
	\frac{X}{2} +\sqrt{ \frac{X^2}{4} + \delta_\sigma^2}
\right) =
|W|\Big(\delta_{\bar\sigma} - \delta_\sigma \Big).
\end{eqnarray}
There is a symmetric solution for this equation
\begin{eqnarray}
\label{symmetric_delta}
\delta_\sigma = \delta_{\bar \sigma} = \sqrt{ 1 - X } = \sqrt{2\nu-1},
    \\
\label{symmetric_nu}
\nu = 1 - X/2.
\end{eqnarray}
Clearly, this solution does not exist when
$X > 1$.
The corresponding free energy difference reads
\begin{eqnarray}
\label{freeSym}
\frac{F_{\rm S}}{V}
=
\frac{F_0}{V}
+
2 N_{\rm F} D_0^2 \int_0^{X} \nu(X')dX'
=
\\
\nonumber
\frac{F_0}{V}
+
2N_{\rm F} D_0^2 \left( X-\frac{X^2}{4} \right),
\end{eqnarray}
which corresponds to
Eq.~(\ref{eq::FSymm}).

\begin{figure}[H]
\centering
\includegraphics[width=0.35\textwidth]{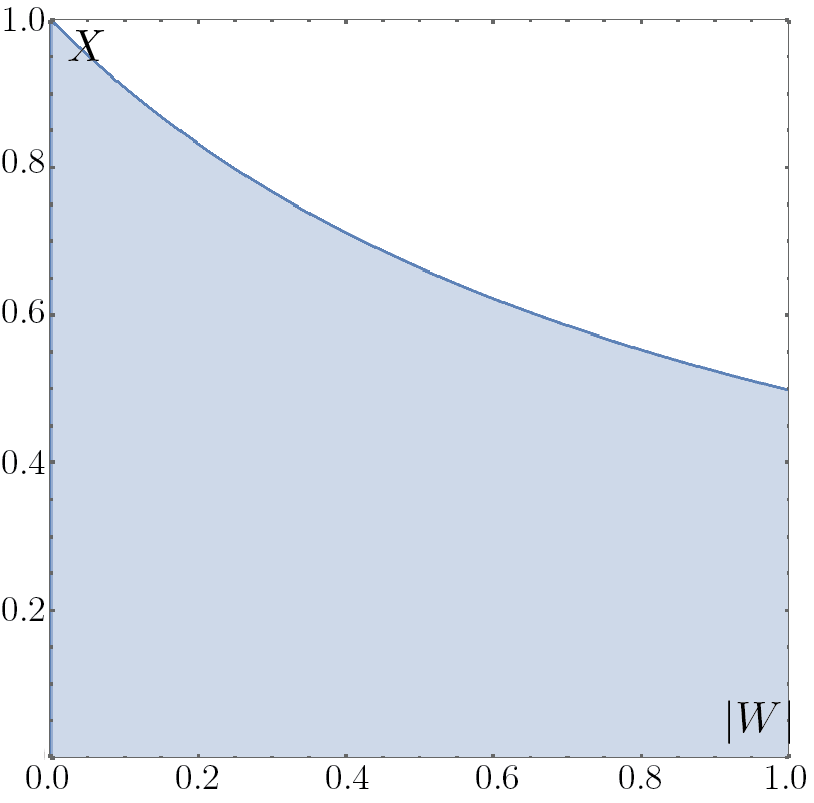}
 \caption{ The region of  existence of the CDW half-metal. The
asymmetric solution of~\eqref{self2}
exists in the shaded (blue) area. When dimensionless doping exceeds the
critical value
$X_c$, 
the consistency relation~(\ref{cons})
is violated, implying the disappearance of the half-metallic phase. The curve
$X_c = X_c (|W|)$ is parametrically determined by Eq.~(\ref{curve}).
Asymptotically, at
$|W|\gg1$,
it becomes a simple hyperbola
$X=|W|^{-1}$.
\label{fig:region}
}
\end{figure}

Next we assume that
sector~$\sigma$
is empty, while
sector~$\bar{\sigma}$
is doped. In terms of dimensionless variables this implies the following
consistency condition
$\delta_{\sigma} > \nu > \delta_{\bar\sigma}$.
It can be expressed, with the help
of~\eqref{connect},
as
\begin{gather}
\label{cons}
\delta_{\sigma} >\sqrt{\delta_{\bar\sigma}^2+X^2}.
\end{gather}
In turn,
Eq.~(\ref{self1})
allows us to derive
\begin{gather}
\label{self2}
\begin{split}
\ln[X +\sqrt{X^2+\delta_{\bar\sigma}^2}]
&=
|W| \Big(\frac{\delta_{\sigma}}{\delta_{\bar\sigma}}-1\Big),
\\
\ln\delta_{\sigma}
&=
|W|\Big(\frac{\delta_{\bar\sigma} }{\delta_{\sigma}}-1\Big).
\end{split}
\end{gather}
This is a system of coupled transcendental equations, which must be solved
with respect to unknown $\delta$'s at fixed $X$.

While we do not know an analytical solution of this system, and ultimately
resort to numerical calculations, some properties of the solution can be
found out analytically. Of particular importance is the
critical value of doping
$X_c = X_c (|W|)$
above which
Eqs.~(\ref{self2})
have no solution consistent with
inequality~(\ref{cons}).
Introducing the parameter $s = \delta_{\bar\sigma}/\delta_{\sigma}$
we immediately obtain the resultant parametric equations
\begin{gather}
    \begin{split}
        X_c(s) &= \frac{\sqrt{1-s^2}}{(1+\sqrt{1-s^2})^{\frac{s}{s+1}}},\\
	|W(s)| &= \frac{s}{1-s^2}\ln[1+\sqrt{1-s^2}].
    \end{split}
    \label{curve}
\end{gather}
“These relations are obtained from the condition that at $X=X_c$ the inequality (92) turned into the equality.
At the
$(|W|, X)$-plane
they specify a curve parameterized by variable
$s \in [0, 1]$.
It is shown in
Fig.~\ref{fig:region},
as a line limiting the area, where the solution exists. We see that, at
$|W|=0$,
sector~$\sigma$ remains empty for the doping level as high as
$X_c (0) = 1$.
For growing $|W|$, however, the dimensionless critical doping
$X_c (|W|)$
decreases.
\begin{figure}[t!]
\centering
\includegraphics[width=0.5\textwidth]{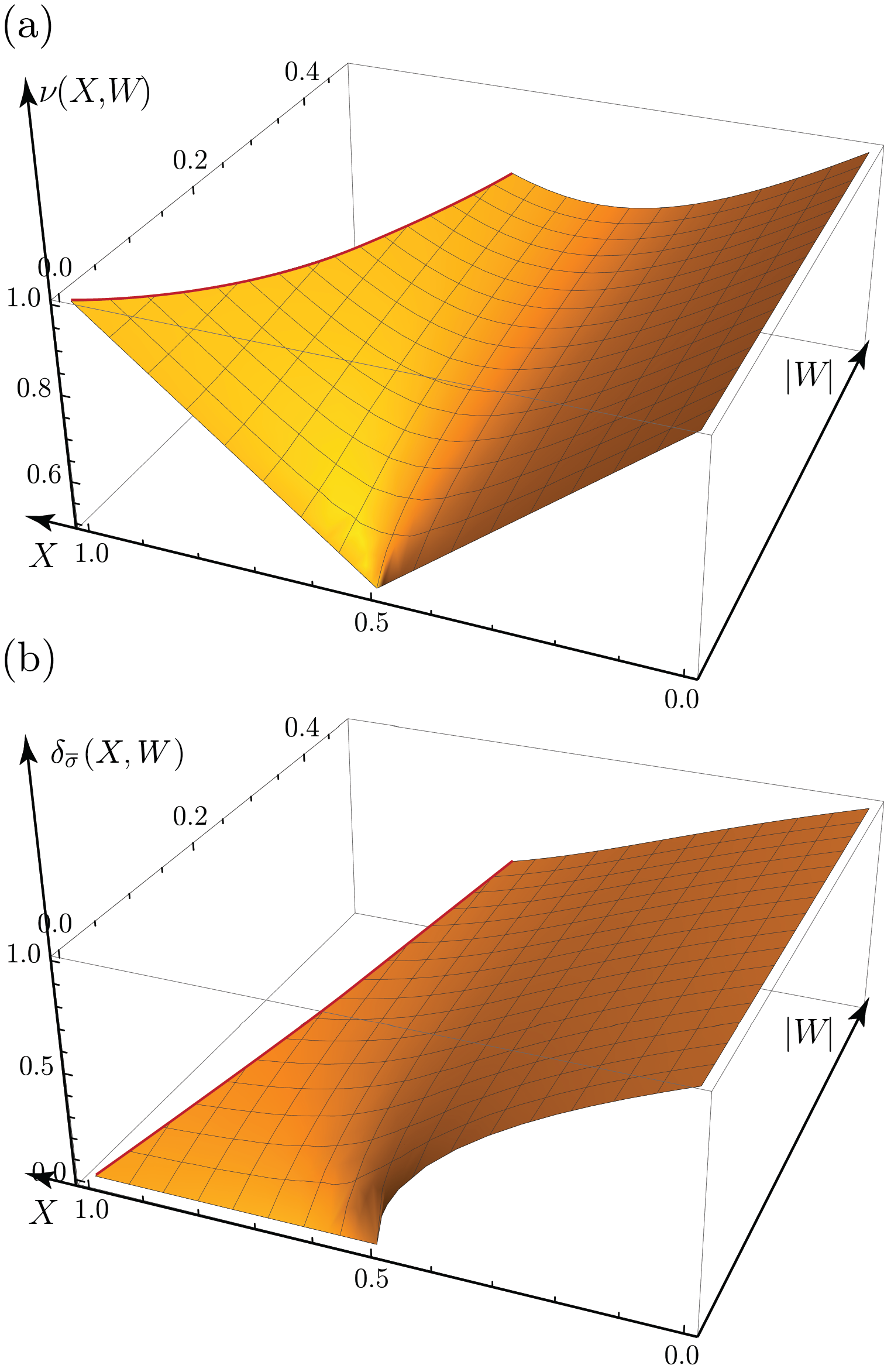}
\caption{(a) Dimensionless chemical potential as a function of dimensionless
particle number $X$ and coupling constant $W$
for the asymmetric CDW state. Only the ${\bar \sigma}$
sector is doped. The  corner-like shape
$\nu = 1/2+|X-1/2|$
is clearly pronounced at zero coupling
$W=0$. (b) Dimensionless order parameter $\delta_{\bar{\sigma}}$ as a function of the same parameters. A simple solution of system ~\eqref{self2} is clearly evident at $W=0$, where  $\delta_{\bar{\sigma}} = \sqrt{1-2X},\ X<1/2$, and the gap closes ($\delta_{\bar{\sigma}} = 0$), at $X>1/2$.
\label{fig:doping}
}
\end{figure}

Solving the self-consistency conditions numerically, we find both
$\delta$'s and $\nu$. Surface in
Fig.~\ref{fig:doping}
shows the behavior of $\nu$ as a function of $X$ and $|W|$. We see, that
at $|W| = 0$,
the $\nu = \nu (X)$ dependence exhibits
a distinct corner shape, as indeed expected.

Having calculated the chemical potential
$\nu(X,W)$,
we may now compute the free energy difference for the asymmetric state
\begin{gather}
\label{asym}
\frac{F_{\rm A}}{V}
=
\frac{F_0}{V}
+
2N_{\rm F} D_0^2 \int_0^X \nu(X',W)dX'.
\end{gather}
This formula is analogous to
Eq.~(\ref{eq::FAsymm}).
Unlike the SDW result, however, we evaluate
Eq.~(\ref{asym})
numerically.

\subsection{Properties of doped CDW phases}

\begin{figure}[t!]
\centering
\includegraphics[width=0.4\textwidth]{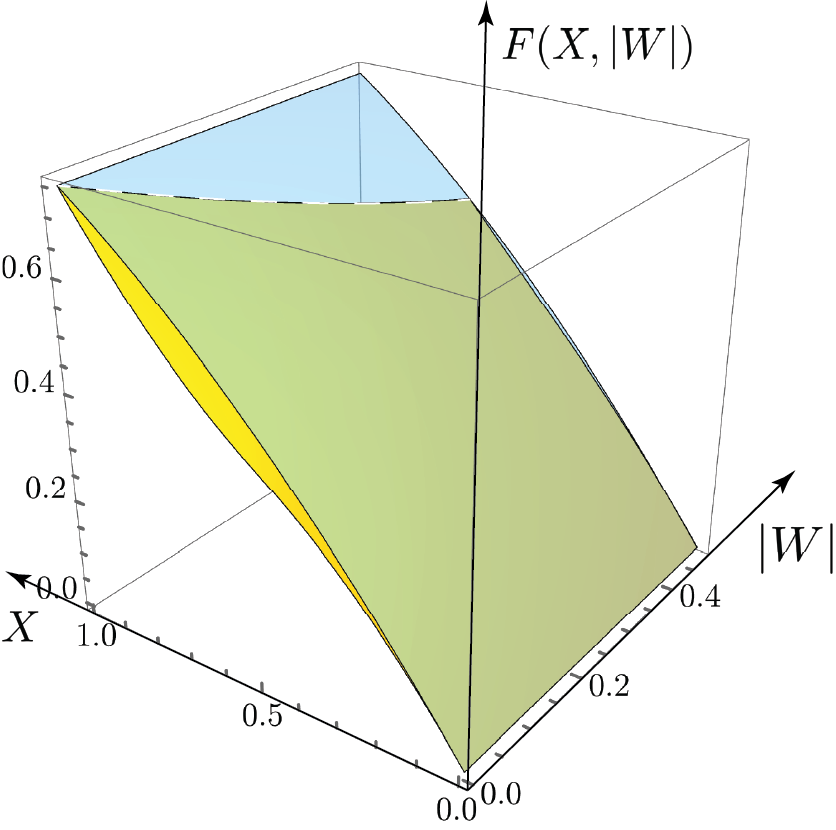}
\caption{Free energy of the symmetric (blue semitransparent surface) and asymmetric (yellow
surface) states of the model as a function of the dimensionless particle number
$X$ and coupling constant $|W|$. One clearly sees that asymmetric phase is
energetically favorable in the entire region, where it exists. Exactly at
the critical doping $X_c(X)$ [the dashed curve, see also Fig.~\ref{fig:region} and Eq.~(\ref{curve})], both the symmetric and asymmetric phases have the same energy. At $X>X_c$, the symmetric phase prevails.
\label{fig:energy}
}
\end{figure}

Let us compare the free energies of the symmetric and asymmetric phases,
see
Eqs.~(\ref{freeSym})
and~\eqref{asym}.
The corresponding plot is presented in
Fig.~\ref{fig:energy}.
Our calculations demonstrate that, if the asymmetric phase exists, it
always has lower energy than the symmetric phase. In other words, for
$X\in [0, X_c(|W|)]$,
the model ground state is the asymmetric phase.

 For higher doping,
the asymmetric phase is impossible. Hence, the curve
$X=X_c(W)$
is the transition line separating two phases (symmetric and asymmetric). Since the doping in a
specific sector changes discontinuously upon transition between symmetric
and asymmetric states, we conclude that the transition is the first order one
(discontinuous).

In the $X_c < X < 1$ range,
the symmetric solution is realized. For even higher doping $X>1$,
the order parameter is completely suppressed, and the system is a metal
with no broken symmetries.

There are several similarities between the symmetric and asymmetric phases.
Both of them are characterized by a finite CDW order parameter. For the
symmetric phase, we can write a concise formula for this order parameter
\begin{eqnarray}
\label{eq::OP_cdw_doped}
|\Delta| = \frac{2 g}{g - 2 \bar{g}_\bot} D_0
	\sqrt{1 - \frac{\gamma x}{2 N_{\rm F} D_0}}.
\end{eqnarray}
For the asymmetric phase, a simple analytical expression valid at arbitrary
$X$ is unknown to us. Further, both phases are metallic, with
well-developed Fermi surfaces, which are two concentric spheres. The radii
of these spheres are determined by
Eq.~(\ref{eq::pocket_momenta}), in which
$\Delta_\sigma$
must be replaced by $D_\sigma$.

Yet, there are important distinctions between the two. In regard to their
metallicity, one must remember that, while the symmetric phase is a kind of
a paramagnetic metal, the asymmetric phase is a half-metal: its
sector~$\sigma$ has no Fermi surface.

In addition, the asymmetry in doping implies finite ferromagnetic polarization:
all fermions introduced by the doping enter sector~$\bar\sigma$, where all
single-fermion states have spin projection~$-\sigma$. Consequently, the total
spin polarization is $-x\sigma/2$. Quite remarkably, the asymmetric phase has a finite
SDW polarization
$\propto
(D_\uparrow - D_\downarrow) \cos ({\bf Q} \cdot {\bf r} + \phi)$.
Thus, the total spin polarization is equal to
\begin{eqnarray}
\label{eq::Sz_cdw_doped}
\langle S_z ({\bf r}) \rangle
=
-x\sigma/2 + 2 g^{-1} (D_\uparrow - D_\downarrow) \cos ({\bf Q}\cdot{\bf r} + \phi),
\end{eqnarray}
Here, $\phi$ is an arbitrary constant determined by the origin in the $\br$  space.
One can check directly that
$\langle S_z ({\bf r}) \rangle \equiv 0$
in the symmetric state.

Let us address the situation, for which the solution of the asymmetric system~\eqref{self2}
disappears. This occurs precisely, when condition~\eqref{cons}
breaks. That is, the particle number attains a
critical value
$X_c=X(W)$
 such that it cannot be
accommodated in a single band pocket (see
Fig.~\ref{fig:region}).
That corresponds to the limiting relation between the gaps
\begin{gather}
    \delta_\sigma^2 = X_c^2 +\delta_{\bar \sigma}^2,\quad \nu = \sqrt{\delta_{\bar \sigma}^2+X_c^2}.
\end{gather}
As doping grows beyond the critical value, the state changes to the symmetric one
$X_\sigma = X_{-\sigma} \equiv X/2$.
This also implies that
\begin{gather}
    \delta_\sigma = \delta_{\bar \sigma},
    \quad
    \nu = \sqrt{\delta_{\bar \sigma}^2+X^2/4}.
\end{gather}
In other words, the gaps and partial dopings in both sectors are identical. The transition between asymmetric and symmetric phases is discontinuous (of the first order).

\begin{figure}[t!]
\centering
\includegraphics[width=0.4\textwidth]{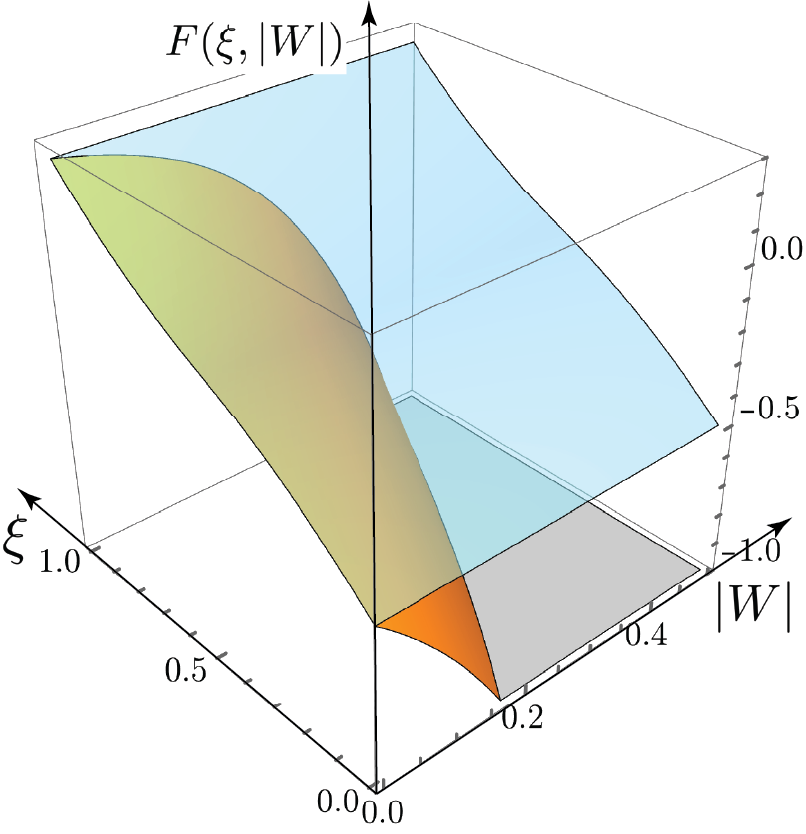}
\caption{The free energy of the CDW (orange) vs SDW (blue) phases for the negative
values of $W$ as a function of dimensionless particle number $\xi$ and
$|W|$. We see that for all relevant parameter values, the CDW state has
the lowest energy.
The degeneracy of the free energy of SDW and CDW at $W = 0$ is pronounced. The gray surface is a clipping plane.
\label{fig:full}
}
\end{figure}

\section{Phase diagram of the model}
\label{sec::phase_diag}

Our analysis has demonstrated that the ground state of the model depends on
doping $x$. Another important parameter affecting the ground state is
$\bar{g}_\bot$:
it controls the switching between the SDW and CDW orders, and, within the
CDW phase, it affects the strength of the order parameter. In this section,
we construct the model's phase diagram that summarizes these trends in a single
layout.

Of course, it is convenient to use dimensionless variables instead of $x$
and
$\bar{g}_\bot$.
For example, we can use
$\bar{g}_\bot/g$
or $W$ as a dimensionless representation of
$\bar{g}_\bot$.
However, dimensionless
$X_\sigma$
is not very suitable for this purpose, since its very definition, see
Eq.~(\ref{eq::Xsigma}),
contains
$\bar g_\bot$.
To circumvent this issue, let us introduce new dimensionless quantities
\begin{eqnarray}
\label{xi}
\xi_\sigma  = \frac{\gamma x_\sigma}{2N_{\rm F}\Delta_{0}},
\quad
\xi = \sum_\sigma \xi_\sigma.
\end{eqnarray}
Note that
$X  = \xi e^{2W}$.

\subsection{SDW order}

To construct the phase diagram, we need to compare the free energies of all
phases we have identified. Let us start with the SDW ordering. We know
that, for low doping,
$x<N_{\rm F} \Delta_0/ \gamma$,
the free energy of the SDW state is expressed by
Eq.~(\ref{eq::FAsymm}).
Exactly at
$x = N_{\rm F} \Delta_0/ \gamma$,
the order parameter
$\Delta_\sigma$
at the sector hosting all doped charge carriers vanishes,
$\Delta_\sigma = 0$,
and remains to be empty for larger $x$.

This nullification of
$\Delta_\sigma$
manifests itself as a second-order transition between two types of the SFHM. Both these
phases correspond to asymmetric state, both are half-metals. The energy of the
high-doping SFHM is
$F_{\rm A} = - N_{\rm F} \Delta_0^2/2 + \gamma^2 x^2/(4 N_{\rm F})$,
for
$x>N_{\rm F} \Delta_0/ \gamma$.
We can express the free energy for the SDW ordering as follows
\begin{gather}
    F_{\rm sdw}(\xi)/V = N_{\rm F }\Delta_0^2
    \begin{cases}
    -1 + 2\xi - \xi^2 ,\quad \xi \leqslant \frac{1}{2},\\
     \xi^2 - \frac{1}{2},\quad \xi>\frac{1}{2}.
    \end{cases}
\end{gather}

\subsection{CDW order}

At
$X < X_c$,
the free energy of the CDW state in the same notation reads
\begin{gather}
F_{\rm cdw}(\xi)/V
=
N_{\rm F}\Delta_0^2e^{-4W}\!
\left[ - 1 \!+\! 2\!\int_0^{\xi e^{2W}}\!\!\! \nu(X)\,dX\right]\!\!.
\end{gather}
This expression must be evaluated numerically.

Let us recall that, at
$X = X_c$,
within the CDW order, there is a phase transition between symmetric and
asymmetric states. The first-order transition line separating these states
is specified by parametric
equations~(\ref{curve}).
In terms of $\xi$ and
$\bar{g}_\bot/g$
variables, we can rewrite these formulas as
\begin{gather}
\label{eq::xi_1}
\begin{split}
    \xi(s) &\equiv \xi_1(s) = \sqrt{1-s^2}(1+\sqrt{1-s^2})^{\frac{s}{1-s}},\\
    \frac{\bar{g}_\bot}{g} &= -\frac{gN_{\rm F}|W(s)|}{1-2gN_{\rm F}|W(s)|}.
\end{split}
\end{gather}
When
$\bar{g} = 0$,
or, equivalently,
$W=0$,
it is trivial to check that
$\xi_1 = 1$.
In the opposite limit, at
$\bar{g}_\bot\gg g$,
we have
$|W| \rightarrow 1/(2 N_{\rm F} g)$
and
$\xi_1 \rightarrow \exp(2|W|)/|W|$.

The symmetric CDW order disappears completely at
$X=1$,
see
Eqs.~(\ref{symmetric_delta})
and~\eqref{symmetric_nu}.
This condition defines a second-order transition line
\begin{gather}
\label{eq::xi_2}
\xi_2(W) =  e^{2|W|}
\end{gather}
that separates a normal metal and a metal with the CDW order parameter. Here, $W$ as a function of $\bar{g}_\bot/g$ is given by Eq.~\eqref{eq::xi_1}.

\begin{figure}[t!]
\centering
\includegraphics[width=0.45\textwidth]{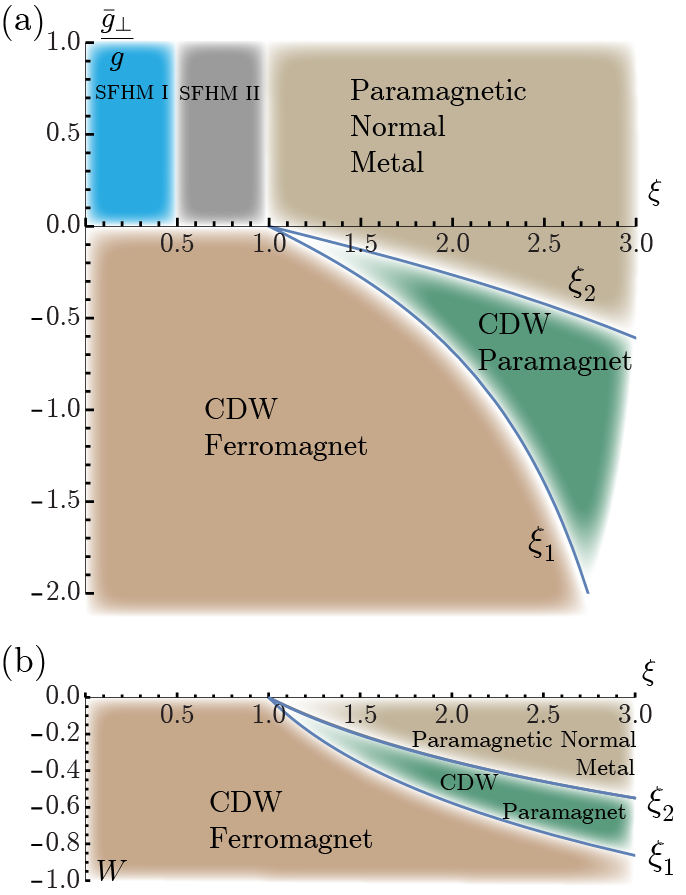}
\caption{Phase diagram.
(a) $\xi = \gamma x/(2N_{\rm F}\Delta_{0})$
is the dimensionless particle number.
Curve
$\xi_1$
($\xi_2$)
is expressed by
Eq.~(\ref{eq::xi_1})
[Eq.~(\ref{eq::xi_2})] We used $gN_{\rm F}=0.5$ for definiteness. Curve $\xi_1$ becomes vertical at $\xi\rightarrow \exp(1/gN_{\rm F})$.
(b) The phase diagram presented in universal units (independent of the coupling parameter $g N_{\rm F}$) $W$ and $\xi$ for negative values of $W$. Here, both curves have exponential asymptotic behavior
$\xi_1(W) \rightarrow \exp(2|W|)/|W|$,
while $\xi_2(W)$ is given by Eq.~\ref{eq::xi_2} at large $W$.
\label{fig:phase}
}
\end{figure}

\subsection{Phase diagram details}

The resultant phase diagram is shown in
Fig.~\ref{fig:phase}.
Comparing
$F_{\rm sdw}$
and
$F_{\rm cdw}$,
we find that, at
$W < 0$,
the CDW always has the lowest energy,
see Fig.~\ref{fig:full}.
As a result, at any negative
$\bar g_\bot$
the CDW phase prevails for any doping.

We see that at
$\bar{g}_\bot < 0$
(at
$\bar{g}_\bot > 0$)
the CDW (the SDW) prevails. This feature can be interpreted as follows:
the effective
$\bar{g}_\bot$
becomes negative only when the charge carriers interact with the lattice, and the
latter is sufficiently soft. Otherwise, the exchange interaction makes the
CDW either metastable, or unstable.

On crossing the line
$\bar{g}_\bot = 0$,
$\xi \in [0, 1]$,
the system demonstrates the first-order transition between the CDW and the
SDW orders. Specifically, the ferromagnetic magnetization experiences a
discontinuity at this transition, as one can prove comparing
Eqs.~(\ref{eq::Sz_sdw_FM})
and~(\ref{eq::Sz_cdw_doped}).

As we have seen in
Sec.~\ref{sec::sdw},
the coupling constant
$\bar{g}_\bot$
does not affect the SDW phase. That is why all phase boundaries in the SDW
phase are vertical straight lines. The $\xi\in[0,1],\ W>0$ strip is split into two halves by $\xi = 0.5$ line separating different types of SFHM phases. Boundary value $\xi = 0.5$ corresponds to a critical doping level $x = N_{\rm F} \Delta_0/\gamma$ [see Eq. (100)] where the order parameter of the doped sector $\delta_{\bar{\sigma}}$ disappears [cf. Eq.~\eqref{minSol1}].

At
$\bar{g}_\bot = 0$,
$\xi = 0.5$,
there is a tricritical point where two SFHM phases meet the CDW
half-metal. Examining this point, one may wonder why there is no second
type of the CDW half-metal, similar to the second type of the SFHM. To
answer this question, let us recall the difference between two SFHM's: the
low-doping SFHM
($0 < \xi < 0.5$)
has finite order parameter in the sector hosting the inserted carriers,
while for the high-doping SFHM
($0.5 < \xi < 1$)
this order parameter is zero. Such a situation is mathematically forbidden
for the CDW half-metal:
Eqs.~(\ref{self2})
have no non-trivial solution for which
$\delta_\sigma = 0$.
In other words, because of the inter-sector coupling in the CDW phase, the
empty-sector order parameter does not allow the nullification of the
filled-sector order parameter. One can treat
$\delta_{\bar \sigma}$
as a kind of external field for
$\delta_\sigma$,
whose presence replaces the second-order transition by a crossover.

At
$\bar{g}_\bot = 0$ and $\xi = 1$,
we see a tetracritical point: the CDW half-metal, the CDW paramagnet, the
SFHM, and normal metal phases coexist there. Of four transition lines
meeting at this point, only one corresponds to the second-order transition:
the CDW paramagnet order parameters smoothly vanish at the transition to
the normal metal, see
Eq.~(\ref{symmetric_delta}).
Three other lines represent discontinuous transitions: for example, one can
check that, on crossing any of these lines, the ferromagnetic magnetization
demonstrates a discontinuity.

\section{Discussion}
\label{sec::discussion}

\subsection{Results and approximations overview}

Our investigation of the model revealed several ordered states that are
sensitive to the doping and interaction structure. We studied magnetic and
metallic properties of these phase and mapped out the phase diagram.

All doped phases on the phase diagram are metallic, with well-developed
Fermi surface. Except for the CDW paramagnet and normal metal, all other
doped phases are ferromagnetic half-metals, with non-trivial spatially
modulated spin textures. The latter circumstance is particularly surprising
in case of the CDW ordered phases: a gappped nonmagnetic insulator acquires
a ferromagnetic polarization, coexisting with a SDW-like spin order, see
Eq.~(\ref{eq::Sz_cdw_doped}).

To assess the validity of these findings, we must also remember the
assumptions that we made in our study, as well as propose possible
enhancements for the model. First of all, all our calculations were
performed at
$T=0$.
Generalization beyond this limit is a matter of the future research.

Further, we employed the mean field approximation. We believe that for a
three-dimensional many-fermion system (and, perhaps, for a two-dimensional
system at
$T=0$) currently, there is
no known mechanism that would qualitatively alter any mean field
conclusion, as long as coupling constants are weak. A brief overview of
recent literature  relevant for this issue can be found in Appendix~A
of  Ref.~\onlinecite{kokanova2021}.

From the very beginning of our investigation, we assumed that no macroscopic
phase separation is allowed: apart from density modulations with the CDW
wave vector
${\bf Q}$,
the density is postulated to be the same everywhere. For this to be true,
the long-range Coulomb interaction must be sufficiently strong. A detailed
exploration of the Coulomb interaction role is an interesting direction for
future studies.

Our density-waves order parameters are characterized by the nesting wave
vector
${\bf Q}$.
In our calculations it was always treated as an invariable quantity.
Nonetheless, it is possible to view
${\bf Q}$
as yet another minimization parameter. Suitable formalism is quite old, see
Ref.~\onlinecite{Rice}.
It is straightforward but cumbersome. At the same time, the energy gain due
to minimization over
${\bf Q}$
is quite insignificant. Thus, we chose to disregard this possibility. As a
justification, one can imagine that
${\bf Q}$
is commensurate with the lattice, thus, its value is pinned by the lattice.
Possible
${\bf Q}$
incommensurability opens yet another direction, in which this research can be
extended: the role of the umklapp interaction.

The electron-lattice coupling is an important element of our model. Thanks
to this interaction, the CDW can win the competition against the SDW phase
(the SDW is typically stronger without the lattice participation due to the
exchange interaction disfavoring the CDW). Since the crystal lattice plays
a rather minor role, it was accounted for within a very minimalist
framework: a single mode, treated as static non-quantum variable
$\vfQ$.
In principle, more realistic description of the lattice may be included
into the model.

\subsection{Stoner mechanism of ferromagnetism}

As we explained in the Introduction, the mechanism discussed in this paper
is unrelated to the Stoner instability. To prove this, let us recall that
the latter implies that a many-fermion system enters a ferromagnetic phase
when the inequality
\begin{eqnarray}
\label{Stoner}
g N_{\rm F} > 1
\end{eqnarray}
becomes valid.
Our paper, on the other hand, is formulated in the framework of the weak coupling. The respective dimensionless small parameter of the analysis is $gN_{\rm F}\ll1$. Therefore, relation ~\eqref{Stoner} is definitely violated.

The weak-coupling limit, which we confined ourselves to, should be seen as
an important theoretical advantage: when coupling constants are small, the
mean field results can be consistently tested with the help of a suitably
designed perturbative expansion. This possibility is to be contrasted with
the Stoner criterion framework. Simplicity and universality
of Eq.~(\ref{Stoner})
creates an impression of an easy-to-use multipurpose tool, applicable for
lattice as well as continuum models. Unfortunately, the reliability of
Eq.~(\ref{Stoner})
as a predictor of a ferromagnetic phase is not beyond reproach. Clearly,
for the Stoner instability to occur, a system must enter intermediate or
strong coupling limit. The arsenal of analytical instruments that are
available in these regimes is severely restricted. Moreover, no controlled
derivation of condition
Eq.~(\ref{Stoner})
is known.

This brings numerical approaches to the forefront of research efforts.
In this regard, there is ample numerical evidence questioning
trustworthiness of condition~(\ref{Stoner}).
Specifically, recent computational
works~\cite{no_FM2009drummond, no_FM2020holzmann, no_FM2022azadi}
indicated that the Stoner ferromagnetism is likely to be {\it never} a ground state
of the jellium model of electron gas. Likewise, the  Monte Carlo
simulations for $^3$He
found no ferromagnetic
phase~\cite{no_FM2015He-3_taddei, no_FM2006He-3_holzmann},
in agreement with experimentally known phase diagram of $^3$He.

Application of the Stoner criterion to a Hubbard-like Hamiltonian may be
inconsistent with numerical data as well, as the authors of
Ref.~\onlinecite{hubbard2022review}
argued. The presence of a ferromagnetic state at the phase diagram of a
triangular lattice model has been discussed in a recent
publication~\cite{morera2024hubbard_3angular},
yet the relevance of the Stoner mechanism to this phase is not immediately
clear.

In the realm of experimental physics, the guidance provided by the
inequality~(\ref{Stoner})
may be quite unreliable as well. Consider, for example, liquid $^3$He.
The interaction between He atoms is typically modeled by the hard-sphere
potential, implying that $g$ in
Eq.~(\ref{Stoner})
is extremely large. Yet, no liquid ferromagnetic phase of $^3$He is known.
Also, we can recall how early optimistic
papers~\cite{vitkalov2001search_for_stoner,shashkin2001ferromag_prl_exper}
reporting ferromagnetism observation in silicon MOSFET were later
criticized~\cite{pudalov2002low}
for data interpretation being too one-sided. Likewise, recent review
paper~\cite{shashkin_kravchenko2019review},
while acknowledging the contribution of
Refs.~\onlinecite{vitkalov2001search_for_stoner,
shashkin2001ferromag_prl_exper}
to the field, did not endorse the Stoner-ferromagnetism interpretation.

Speaking more broadly, the Stoner proposal ignores various correlation
effects that affect the energies of both polarized and paramagnetic
liquids. In lattice models, with their assortment of ordered states,
competition with other phases may additionally impede ferromagnetism
stabilization. (To appreciate the latter issue, one may examine phase
diagrams in
Ref.~\onlinecite{Igoshev2015hubbard}
that compare blanket Stoner criterion prediction with rich phase diagrams
obtained within either slave-boson or Hartree--Fock approximations.) These
rather common features of many-fermion models make simple and universal
prescription~(\ref{Stoner})
quite unreliable in many important situations. Therefore, while the
Stoner criterion remains a popular research tool even today, the
associated issues must not be overlooked.

Despite these problems, it is difficult to ignore the fact that the
Bloch--Stoner calculations were one of the first quantum-mechanical
many-body theories, a pioneering attempt to derive a consistent explanation
of an ordered many-electron state in condensed matter. Ultimately, it is
hardly a surprise that over many decades of research our understanding of
solid state magnetism has evolved, motivating re-examination of previous
concepts.

\subsection{Possible realizations of the model: graphene multilayers}

In connection with possible realization of our model in experiment, it is
necessary to remember that the ingredients required by our proposal are not
that unique. There are, indeed, substantial variety of materials
demonstrating nesting-driven SDW or CDW phases (although we assumed
spherical Fermi surfaces, this is mostly a matter of convenience, and any
type of nested Fermi surface suits us). If these can be doped, the
stabilization of a ferromagnetic state becomes an option.

 In this connection, it is appropriate to mention hexaboride compounds.
They typically have the nested structure of a Fermi surface (see ~[\onlinecite{Souma2003}] for CaB$_6$ and \onlinecite{delley2004properties},~\onlinecite{ishizawa1980transverse} for LaB$_6$).
Crucially, the ferromagnetism attributed to the discussed mechanism was indeed reported to be observed in CaB$_6$~\cite{young1999high}. We would also like to mention the so-called fractional metal states in graphene-based materials. The fractional metals can be
conceptualized as a generalization of ferromagnetic phase for a system with
both spin and valley degeneracies. These states have been theoretically
predicted in
Refs.~\onlinecite{fracmet2021prblett, AB_quarter_metal}.
Experimental observations have been reported in
Ref.~\onlinecite{trilayer2021FraM_nature, zhou2022isospin}.

To establish their relevance for our model, let us observe that
graphene bilayers demonstrate an instability toward an insulating ordered
state~\cite{bilayer_review2016}.
The instability is driven either by nesting (for AA-stacking), or by
quadratic band touching (for AB-staking). Depending on details, doped
system may demonstrate either spatially inhomogeneous state, or, instead,
polarization in terms of valley and spin-related indices, similar to the
behavior discussed in
subsection~\ref{subsec::sdw_doped}.
Additional details can be found in
Refs.~\onlinecite{fracmet2021prblett, AB_quarter_metal}.


\section{Conclusions}

Using the mean field approximation, we investigated a model with electrons and
holes whose Fermi surfaces are perfectly nested. It was assumed that the
fermions interact with each other and with the lattice. All interactions
are weak. To suppress inhomogeneous states, strong long-range Coulomb
repulsion was included into the model. When undoped, the ground state of
such a model is insulating and demonstrates a density-wave order, either SDW,
or CDW. The doping weakens both types of order. Additionally, finite
ferromagnetic polarization emerges. We constructed the phase diagram of the
model, and studied various properties of the ordered phases.

\section*{Acknowledgements}
We are grateful to Alexander Chernyshev
for timely drawing our attention to previous work on the subject.

\section*{Data availability}
The data that support the findings of this study are available
from the corresponding author on reasonable request.

\bibliographystyle{apsrevlong_no_issn_url}

\bibliography{bibliography}

\end{document}